\documentclass[a4paper,11pt]{article}

\usepackage[utf8]{inputenc}
\usepackage{etex}

\usepackage{xspace} 

\usepackage{jheppub}

\usepackage{feynmp}

\allowdisplaybreaks

\newcommand{\arxivversion}{1} 

\usepackage[usenames,dvipsnames]{xcolor}	

\usepackage{mathrsfs}
\usepackage{textcomp}

\usepackage{{amsmath,amsfonts,latexsym, amstext, amssymb, amsthm}}
\usepackage[english]{babel}
\usepackage{braket}
\usepackage{dsfont}
\usepackage{tabularx}
\usepackage{longtable}
\usepackage{comment}
\usepackage{pstricks}
\usepackage{pst-node}
\usepackage{pst-plot}

\usepackage{multirow}

\usepackage{longtable}
\usepackage{multirow}
\usepackage{pdflscape}
\usepackage[textsize=footnotesize]{todonotes}
\usepackage{mathtools}

\newcommand{\RR}{\ensuremath{\mathbb{R}}}

\newcommand{\NN}{\ensuremath{\mathbb{N}}}

\newcommand{\G}{\ensuremath{G}\xspace}
\newcommand{\UN}{\ensuremath{\text{U}(N)}\xspace}
\newcommand{\SUN}{\ensuremath{\text{SU}(N)}\xspace}
\newcommand{\U}[1]{\ensuremath{\text{U}(#1)}\xspace}
\newcommand{\SU}[1]{\ensuremath{\text{SU}(#1)}\xspace}

\newcommand{\PSLs}[2]{\ensuremath{\text{PSL}(#1|#2)}\xspace}

\newcommand{\su}[1]{\ensuremath{\mathfrak{su}(#1)}\xspace}
\newcommand{\so}[1]{\ensuremath{\mathfrak{so}(#1)}\xspace}

\newcommand{\complexi}{i}

\newcommand{\EulerPhi}{\varphi}

\newcommand{\YM}{{\mathrm{\scriptscriptstyle YM}}}
\newcommand{\maxset}[1]{\max{\{#1\}}}

\DeclareMathOperator{\tr}{tr}

\DeclareMathOperator{\phaneq}{\phantom{{}=}}
\newcommand{\phaneqtimes}{\qquad \times}

\newcommand{\colors}{s}

\newcommand{\e}{\operatorname{e}}
\newcommand{\cstar}{\ensuremath{\ast}}
\newcommand{\numberdot}[1]{#1}

\newcommand{\ev}[1]{\langle #1 \rangle}

\newcommand{\comm}[2]{[#1,#2]}
\newcommand{\acomm}[2]{\{#1,#2\}}
\newcommand{\starcomm}[2]{\comm{#1}{#2}_\cstar}
\newcommand{\staracomm}[2]{\acomm{#1}{#2}_\cstar}

\newcommand{\vac}{\operatorname{\mid}0 \operatorname{\rangle}}

\newcommand{\eqndot}{\, .}
\newcommand{\eqncom}{\, ,}

\newcommand{\nnl}{\nonumber \\}

\newcommand{\de}{\operatorname{d}\!}
\DeclareMathOperator{\dop}{d}
\newcommand{\measure}[1]{\dop\! #1}
\newcommand{\Diff}[2]{\frac{\dop^{#2}}{\dop\!{#1}^{#2}}}
\newcommand{\IntOp}[3]{\int^{#3}_{#2} \!\!\! \dop\! #1 \,}

\newcommand{\cA}{\mathcal{A}}

\newcommand{\cF}{\mathcal{F}}

\newcommand{\cN}{\mathcal{N}}
\newcommand{\cO}{\mathcal{O}}
\newcommand{\cP}{\mathcal{P}}

\newcommand{\cZ}{\mathcal{Z}}

\newcommand{\ba}{\mathbf{a}}
\newcommand{\bb}{\mathbf{b}}
\newcommand{\bc}{\mathbf{c}}
\newcommand{\bq}{\mathbf{q}}

\newcommand{\alphadot}{{\dot{\alpha}}}

\newcommand{\thooft}{'t~Hooft\xspace} 

\renewcommand{\H}{{\text{H}}}

\newcommand{\AdSCFTc}{AdS/CFT correspondence\xspace}
\newcommand{\Polya}{P\'{o}lya\xspace}

\newcommand{\RxSt}{\ensuremath{\mathbb{R}\times \text{S}^3}\xspace}
\newcommand{\Nfour}{$\mathcal{N}=4$\xspace}
\newcommand{\NfSYM}{\Nfour SYM\xspace}
\newcommand{\NfSYMt}{\Nfour SYM theory\xspace}

\newcommand{\gidef}{${\gamma}_{i}$-deformation\xspace}
\newcommand{\gidefd}{${\gamma}_{i}$-de\-formed\xspace}
\newcommand{\expectationvalue}{expectation value\xspace}

\newcommand{\ferm}{\psi}
\newcommand{\antiferm}{\bar{\psi}}

\newcommand{\sufphi}{\varphi}
\newcommand{\cfstrength}{\cF}
\newcommand{\cantifstrength}{\bar{\cfstrength}}

\newcommand{\cder}{D}

\newcommand{\aosc}{\ba}
\newcommand{\aoscdag}{\aosc^\dagger}
\newcommand{\bosc}{\bb}
\newcommand{\boscdag}{\bosc^\dagger}
\newcommand{\cosc}{\bc}

\newcommand{\akind}[1][ ]{{a^{#1}}}
\newcommand{\bkind}[1][ ]{{b^{#1}}}
\newcommand{\ckind}[1][ ]{{c^{#1}}}

\newcommand{\akindsite}[2][ ]{a^{#1}_{(#2)}}
\newcommand{\bkindsite}[2][ ]{b^{#1}_{(#2)}}
\newcommand{\ckindsite}[2][ ]{c^{#1}_{(#2)}}

\newcommand{\centralchargeopdensity}{\ensuremath{C}}

\DeclareMathOperator{\T}{T}

\newlength{\eqoff}
\DeclareMathOperator{\Kop}{K}

\DeclareMathOperator{\D}{D}

\newcommand{\atkind}[1][ ]{{\tilde{a}^{#1}}}
\newcommand{\btkind}[1][ ]{{\tilde{b}^{#1}}}
\newcommand{\ctkind}[1][ ]{{\tilde{c}^{#1}}}

\newcommand{\feynmpdefinitionsdiags}{

\fmfcmd{%
thin := 1pt; 
thick := 2thin;
arrow_len := 4mm;
arrow_ang := 15;
curly_len := 3mm;
dash_len := 1.8mm; 
dot_len := 1.5mm; 
wiggly_len := 2mm; 
wiggly_slope := 60;
zigzag_len := 2mm;
zigzag_width := 2thick;
decor_size := 5mm;
dot_size := 2thick;
}

\fmfcmd{%
style_def plain_sarrow expr p =
  cdraw p;
  shrink (0.55); 
  cfill (arrow p);
  endshrink;
enddef;
style_def dashes_sarrow expr p =
  draw_dashes p;
  shrink (0.55);
  cfill (arrow p);
  endshrink;
enddef;
style_def plain_srarrow expr p =
  cdraw p;
  shrink (0.55);
  cfill (arrow (reverse p));
  endshrink;
enddef;
style_def dashes_srarrow expr p =
  draw_dashes p;
  shrink (0.55);
  cfill (arrow (reverse p));
  endshrink;
enddef;
marksize=2mm;
def draw_mark(expr p,a) =
  begingroup
    save t,tip,dma,dmb; pair tip,dma,dmb;
    t=arctime a of p;
    tip =marksize*unitvector direction t of p;
    dma =marksize*unitvector direction t of p rotated -45;
    dmb =marksize*unitvector direction t of p rotated 45;
    linejoin:=beveled;
    draw (-.5dma.. .5tip-- -.5dmb) shifted point t of p;
  endgroup
enddef;
style_def derplain expr p =
    save amid;
    amid=.5*arclength p;
    draw_mark(p, amid);
    draw p;
enddef;
}
}

\newcommand{\aswfonelabelf}[6][]{%
\settoheight{\eqoff}{$\times$}%
\setlength{\eqoff}{0.5\eqoff}%
\addtolength{\eqoff}{-7.5\unitlength}%
\raisebox{\eqoff}{%
\fmfframe(5,0)(5,0){%
\begin{fmfchar*}(20,15)
\fmfleft{v1}
\fmfright{v2}
\fmffixed{(0.50w,0)}{vc1,vc2}
\fmf{#2}{v1,vc1}
\fmf{#2}{vc2,v2}
\fmf{#3}{vc1,vc2}
\fmf{#4}{vc2,vc1}
\fmfcmd{pair verta, vertb; verta = vloc(__v1); vertb = vloc(__v2);}
\fmfiv{label=#5,l.a=180,l.dist=0.04w}{verta}
\fmfiv{label=#6,l.a=0,l.dist=0.04w}{vertb}
\fmffreeze
\fmfposition
\fmfipath{p[]}
\fmfipair{vm[]}
\fmfiset{p1}{vpath(__v1,__vc1)}
\fmfiset{p2}{vpath(__vc1,__vc2)}
\fmfiset{p3}{reverse vpath(__vc2,__vc1) }
\fmfiset{p4}{vpath(__vc2,__v2)}
{#1}
\end{fmfchar*}}}}

\numberwithin{equation}{section}

\usepackage{etex}
\makeatletter
\DeclareRobustCommand*{\bfseries}{%
  \not@math@alphabet\bfseries\mathbf
  \fontseries\bfdefault\selectfont
  \boldmath
}
\makeatother

\usepackage{lipsum}

\ifnum\arxivversion=1
  \makeatletter
  \if@titlepage
    \renewenvironment{abstract}{%
        \titlepage
        \null\vfil
        \@beginparpenalty\@lowpenalty
        \begin{center}%
          \bfseries \abstractname
          \@endparpenalty\@M
        \end{center}}%
       {\par\vfil\null\endtitlepage}
  \else
    \renewenvironment{abstract}{%
        \if@twocolumn
          \section*{\abstractname}%
        \else
          \small
          \begin{center}%
            {\bfseries \abstractname\vspace{-.5em}\vspace{\z@}}%
          \end{center}%
          \quotation
        \fi}
        {\if@twocolumn\else\endquotation\fi}
  \fi
  \makeatother

\else
\fi

\title{One-Loop Partition Functions in \\Deformed $\mathcal{N}=4$ SYM Theory}
\author{Jan Fokken and Matthias Wilhelm}

\ifnum\arxivversion=1
\else
\affiliation{Institut für Mathematik und Institut für Physik,
Humboldt-Universität zu Berlin\\
IRIS Gebäude, 
Zum Großen Windkanal 6, 
12489 Berlin}
\emailAdd{fokken@physik.hu-berlin.de}
\emailAdd{mwilhelm@physik.hu-berlin.de}
 \abstract{We study the thermodynamic behaviour of the real $\beta$- and $\gamma_i$-deformation of $\mathcal{N}=4$ Super Yang-Mills theory on \RxSt in the planar limit. These theories were shown to be the most general asymptotically integrable supersymmetric and non-supersymmetric field-theory deformations of $\mathcal{N}=4$ Super Yang-Mills theory, respectively. We calculate the first loop correction to their partition functions using an extension of the dila\-ta\-tion-operator and \Polya-counting approach. In particular, we account for the one-loop finite-size effects which occur for operators of length one and two. Remarkably, we find that the $\cO(\lambda)$ correction to the Hagedorn temperature is independent of the deformation parameters, although the partition function depends on them in a non-trivial way.}
\arxivnumber{1411.7695}
\fi

\begin{document}


\ifnum\arxivversion=1

\begingroup\parindent0pt
\begin{flushright}\footnotesize
\texttt{HU-MATH-2014-34}\\
\texttt{HU-EP-14/56}
\end{flushright}
\vspace*{4em}
\centering
\begingroup\LARGE
\bf
One-Loop Partition Functions in \\
Deformed $\mathcal{N}=4$ SYM Theory%
\par\endgroup
\vspace{2.5em}
\begingroup\large
{\bf Jan Fokken,
Matthias Wilhelm}
\par\endgroup
\vspace{1em}
\begingroup\itshape
Institut f\"ur Mathematik und Institut f\"ur Physik\\
Humboldt-Universit\"at zu Berlin\\
IRIS Geb\"aude \\
Zum Gro\ss en Windkanal 6 \\
12489 Berlin
\par\endgroup
\vspace{1em}
\begingroup\ttfamily
\{fokken, mwilhelm\}@physik.hu-berlin.de \\
\par\endgroup
\vspace{2.5em}
\endgroup
\thispagestyle{empty}

\begin{abstract}
We study the thermodynamic behaviour of the real $\beta$- and $\gamma_i$-deformation of $\mathcal{N}=4$ Super Yang-Mills theory on \RxSt in the planar limit. These theories were shown to be the most general asymptotically integrable supersymmetric and non-supersymmetric field-theory deformations of $\mathcal{N}=4$ Super Yang-Mills theory, respectively. We calculate the first loop correction to their partition functions using an extension of the dila\-ta\-tion-operator and \Polya-counting approach. In particular, we account for the one-loop finite-size effects which occur for operators of length one and two. Remarkably, we find that the $\cO(\lambda)$ correction to the Hagedorn temperature is independent of the deformation parameters, although the partition function depends on them in a non-trivial way.
\end{abstract}


\newpage

\setcounter{tocdepth}{2}
\par\noindent\hrulefill\par\vskip 0em
\tableofcontents
\par\noindent\hrulefill\par\vskip-4.3em

\else

\maketitle

\fi






\section{Introduction}
It is a long-standing problem in gauge theories to understand the phase transition between the weakly and strongly coupled regimes; e.g.\ in QCD it corresponds to the confinement-deconfinement transition. In this paper, we address the analogous issue for a class of (asymptotically) integrable deformations of \NfSYMt on \RxSt.

The maximally supersymmetric \NfSYMt has seen impressive advances during the last one and a half decades, in particular through the $\text{AdS}/\text{CFT}$ correspondence \cite{Maldacena:1997re,Gubser:1998bc,Witten:1998qj} and integrability in the 't Hooft limit; see \cite{Beisert:2010jr} for a review on the latter. This has also spurred interest in less (super) symmetric deformations of this theory that share the properties of integrability and a dual string theory description; see \cite{Zoubos:2010kh,vanTongeren:2013gva} for reviews. The prime example for continuous deformations of \NfSYMt is the (real) $\beta$-deformation, a special case of the $\cN=1$ exactly marginal deformations which were classified by Leigh and Strassler \cite{Leigh:1995ep}. It can be obtained by applying a Moyal-like $\ast$-product to the ($\cN=1$ superspace) action of \NfSYMt. This $\ast$-product depends on the three Cartan charges of the original $\SU{4}$ R-symmetry and the real parameter $\beta$. By generalising this $\ast$-product, the non-supersymmetric $\gamma_i$-deformation was proposed \cite{Frolov:2005dj}, which depends on three real parameters. The $\beta$- and $\gamma_i$-deformation are respectively the most general $\cN=1$ supersymmetric and non-supersymmetric, asymptotically integrable field-theory deformations of \NfSYMt \cite{Beisert:2005if}. Their conjectured string-theory duals can be found by applying three TsT-transformations, i.e.\ sequences of T-duality, coordinate shift and T-duality, to the $\text{S}^5$ factor of the $\text{AdS}_5\times\text{S}^5$ background of type IIB string theory \cite{Lunin:2005jy,Frolov:2005dj}.\footnote{We will address important subtleties in these statements as well as recent developments later.}

A convenient way to study the thermodynamic behaviour of gauge theories is via the thermal partition function on \RxSt:
\begin{equation}\label{eq: partition function intro}
 \cZ(T)=\tr_{\RxSt}[e^{- H/T}]\eqncom
\end{equation}
where $H$ denotes the Hamilton operator, $T$ the temperature in units of the Boltzmann constant and the trace is taken over all states on $\RxSt$. In confining theories, the radius $R$ of the three-sphere serves as an effective IR cutoff that stops the running of the coupling constant; tuning it makes the phase transition accessible to perturbation theory. Hence, this setup allows to study confining and non-confining theories on the same footing. In order to observe a sharp phase transition on the compact space $\text{S}^3$, the 't Hooft limit has to be taken. Below the critical temperature $T_\text{H}$, the partition function \eqref{eq: partition function intro} is independent of the number of colours $N$. Above $T_\text{H}$, it scales as $N^2$ and thus diverges \cite{Witten:1998qj}.\footnote{The critical temperature is also called Hagedorn temperature, named after Rolf Hagedorn, who studied a phase transition in the strong force even before its description in terms of QCD was established \cite{Hagedorn65}.} The partition function \eqref{eq: partition function intro} and its phase structure were investigated for several theories via a direct path integral approach on \RxSt \cite{AMMPR03,AMMPR05,AMR06,Mussel:2009uw}.

For a conformal field theory like \NfSYMt, a conformal mapping from \RxSt to $\RR^4$ can be used in order to express \eqref{eq: partition function intro} as
\begin{equation}\label{eq: partition function intro D}
 \cZ(T)=\tr_{\RR^4}\left[x^D\right]\eqncom
\end{equation}
where $x=e^{- 1/RT}$ and $D$ is the dilatation operator \cite{Witten:1998qj}. In comparison to the path integral approach, this saves one loop order in each calculation.\footnote{For instance, the contribution with one momentum loop in the path-integral approach can be obtained from the classical, i.e.\ zero-loop, dilatation operator. In this paper, we follow the counting based on the dilatation-operator approach.} The problem to sum over all states on $\mathbb{R}^4$, i.e.\ gauge-invariant composite operators, then reduces to the enumeration of all graded cyclic spin-chains or necklaces. In free \NfSYMt, this problem was solved by the means of \Polya theory in \cite{Sundborg:1999ue}.\footnote{Note the slight mistake in \cite{Sundborg:1999ue} with respect to the grading.} A central role in this method is played by the single-site partition function
\begin{equation}
 z(x)=\tr_{\cA}\left[x^{\mathfrak{D}_0}\right]\eqncom
\end{equation}
where $\mathfrak{D}_0$ denotes the classical dilatation-operator density and $\cA$ is the alphabet of single-site states from which all single-trace operators can be built. The first loop correction to the \NfSYMt result was calculated via an extension of \Polya theory in \cite{Spradlin:2004pp}. In addition to $z(x)$, it employs the two generalised expectation values
\begin{align}
 \label{eq: SV 3.10 intro}
 \ev{ \mathfrak{D}_2(x) } &= \tr_{\cA\times\cA}\left[x^{\mathfrak{D}_0}\mathfrak{D}_2\right] \eqncom \\
 \label{eq: SV 3.11 intro}
 \ev{ P\mathfrak{D}_2(w,y) }&=\tr_{\cA\times\cA}\left[\cP w^{\mathfrak{D}_0}y^{\mathfrak{D}_0}\mathfrak{D}_2\right] \eqncom 
\end{align}
where $\mathfrak{D}_2$ is the one-loop dilatation-operator density of \cite{Beisert:2003jj} and $\cP$ is the graded permutation operator.

The free result of \cite{Sundborg:1999ue} is also valid for the $\beta$- and $\gamma_i$-deformation, as the $\ast$-product only affects the interactions. However, if we want to employ the one-loop method of \cite{Spradlin:2004pp}, we have to face several important subtleties occurring in the deformed theories that were recently investigated in \cite{Fokken:2013aea,Fokken:2013mza,Fokken:2014soa}.

The $\beta$-deformation is only conformally invariant if the gauge group is chosen to be \SUN. For gauge group \UN, quantum corrections introduce a running double-trace coupling in the component action, which breaks conformal invariance. This coupling is at its non-vanishing fix-point value in the \SUN theory \cite{Hollowood:2004ek}. For the purpose of our one-loop calculation, both theories are sufficiently conformally invariant, as the effects of nonconformality only set in at higher loop orders.  However, note that at higher loop orders the computation of \eqref{eq: partition function intro} via \eqref{eq: partition function intro D} is only valid for the conformal \SUN theory. Moreover, the asymptotic one-loop dilatation operator of \cite{Beisert:2005if} acquires finite-size corrections, i.e.\ an explicit dependence on the length of the spin chain. These effects were intensively studied in \cite{Fokken:2013mza} and at one loop occur for spin chains of length one and two for gauge group \UN and \SUN, respectively. For gauge group \SUN, they can be traced back to the double-trace structure of the propagator and the aforementioned double-trace coupling, whereas they are due to the well-known wrapping effect \cite{Serban:2004jf,Beisert:2004hm,Sieg:2005kd} for gauge group \UN.

The $\gamma_i$-deformation, on the other hand, is not conformally invariant, neither for gauge group \UN nor \SUN \cite{Fokken:2013aea}, at least not for the candidate proposed in \cite{Frolov:2005dj} and all its natural Lagrangian extensions. Here, quantum corrections induce running double-trace couplings without a fix point. This poses very interesting and up to now unsolved questions in the context of the $\text{AdS}/\text{CFT}$ correspondence; see \cite{Fokken:2013aea} for a discussion of different possibilities. For the purpose of our one-loop calculation in this paper, the nonconformality of the $\gamma_i$-deformation itself is no problem. However, a problem arises concerning the uniqueness of the theory: there are many theories that share the single-trace structure of the action proposed in \cite{Frolov:2005dj} and hence the planar asymptotic dilatation operator of \cite{Beisert:2005if}.\footnote{See the discussion in \cite{Fokken:2013aea}.} They differ only in the multi-trace structure and thus in finite-size effects. For the sake of definiteness, we will focus on the candidate proposed in \cite{Frolov:2005dj} with gauge group \UN here. We will see later that the temperature of the phase transition is indeed independent of spin chains of small lengths and thus finite-size effects; it is entirely driven by spin chains of large lengths.

This paper is structured as follows.
In section \ref{sec:deformed_N4_SYM}, we introduce the $\beta$- and $\gamma_i$-de\-for\-ma\-tion; in particular we provide the one-loop dilatation operator.
In section \ref{sec:Partition functions via Polya}, we give a short summary of the method of \cite{Spradlin:2004pp} and modify it to account for finite-size effects.
In section \ref{sec:ingredients}, we compute the necessary ingredients for the partition function: $z(x)$, $\ev{P\mathfrak{D}_2^{L\geq 3}(w,y)}$, $\ev{\mathfrak{D}^{L\geq 3}_2(x)}$, as well as the finite-size correction term $Z^{(1)}_{\text{f.s.c.}}(x)$.
In section \ref{sec:Hagedorn temperature}, we discuss the resulting partition function. In particular, we compute the temperature of the phase transition up to and including the first loop order.
Section \ref{sec:results} contains our conclusion and outlook.
We provide several appendices. 
In appendix \ref{app:fermionic self-energies}, we calculate the one-loop anomalous dimensions of the fermionic $L=1$ operators, which are required in section \ref{sec:deformed_N4_SYM}.
In appendix \ref{sec:Fields, states and the oscillator picture}, we give details on our conventions concerning the spin-chain picture and the required matrix elements of the one-loop dilatation operator of \NfSYMt.
Appendices \ref{app: PD2 calculation}--\ref{app: summation identities} contain details on the calculations of section \ref{sec:ingredients}.
In appendix \ref{app: path integral}, we provide a check of our results for gauge group \UN via a modification of the method and calculation of \cite{Mussel:2009uw}.

\section{\texorpdfstring{Deformed \NfSYM theory}{Deformed N=4 SYM theory}}\label{sec:deformed_N4_SYM}
In this section, we introduce the $\beta$- and $\gamma_i$-deformation and give the dilatation operator up to the first loop order.

Both deformed gauge theories are closely related to their parent \NfSYMt. The single-trace part of their respective actions inherits the properties of \NfSYMt as shown in \cite{Mauri:2005pa,Ananth:2006ac,AKS07}.\footnote{Note that the properties of multi-trace parts of the actions are not captured by the arguments in \cite{Mauri:2005pa,Ananth:2006ac,AKS07}.} They can be obtained by replacing all products in the action of the undeformed \NfSYMt by Moyal-like $\ast$-products. For two field $A$ and $B$, the $\ast$-product is defined as
\begin{equation}\label{eq:starproduct}
 A\cstar B= A B \e^{\frac{\complexi}{2}\mathbf{q}_{A}\wedge\mathbf{q}_{B}} \eqncom
\end{equation}
where $\mathbf{q}_{A}=(q_A^1,q_A^2,q_A^3)$ and $\mathbf{q}_{B}=(q_B^1,q_B^2,q_B^3)$ are the charge vectors of the fields associated with the Cartan subgroup of the $\SU{4}_{\text{R}}$ symmetry group of the undeformed theory, see table \ref{tab: su(4) charges}.
\begin{table}[hb]
\centering
$\begin{array}{l|c|ccc|ccccc}
	B&D_{\alpha\dot{\beta}}&\phi^1&\phi^2&\phi^3 & \psi^1_{\alpha}&\psi^2_{\alpha}&\psi^3_{\alpha}&\psi^4_{\alpha}\\
	& & & & & & & &\\[-0.4cm]
	\hline
	& & & & & & & &\\[-0.4cm]
	q^1_B & 0 & 1 & 0 & 0 &  +\frac12 & -\frac12 & -\frac12 & +\frac12 \\
	& & & & & & & &\\[-0.4cm]
	q^2_B &  0 & 0 & 1 & 0 &-\frac12 & +\frac12 & -\frac12 & +\frac12\\
	& & & & & & & &\\[-0.4cm]
	q^3_B &  0 & 0 & 0 & 1 &-\frac12 & -\frac12 & +\frac12 & +\frac12\\
\end{array}$
\caption{$\SU4_{\text{R}}$ Cartan charges of the fields of \NfSYMt \cite{Beisert:2005if}. The respective anti-fields carry the opposite charges.}\label{tab: su(4) charges}
\end{table}
The antisymmetric product of the charge vectors is given by 
\begin{equation}\label{eq: antisymmetric product}
\mathbf{q}_A\wedge \mathbf{q}_B=(\mathbf{q}_A)^{T}\mathbf{C}\,\mathbf{q}_B
\eqncom \qquad
\mathbf{C}=\begin{pmatrix}
0 & -\gamma_3 & \gamma_2 \\
\gamma_3 & 0 & -\gamma_1 \\
-\gamma_2 & \gamma_1 & 0 
\end{pmatrix}
\eqndot
\end{equation} 
In concrete applications, it can be conveniently written in terms of the following linear combinations of the deformation parameters: 
\begin{equation}
\gamma^\pm_i=\pm\frac 12 (\gamma_{i+1}\pm\gamma_{i+2})\eqncom
\end{equation}
where cyclic identification $i+3\sim i$ is understood.

The single-trace part of the Euclidean action of the $\beta$- and $\gamma_i$-deformation can be given as
\begin{equation}\label{eq: component action single trace}
\begin{aligned}
S_{\text{s.t.}} &=\int\de^4x\,\tr\Big(
 -\frac{1}{4} F^{\mu\nu}F_{\mu\nu}- (\D^{\mu}\bar\phi_j)\D_{\mu}\phi^j
 +i \bar\psi^{\dot\alpha}_A \D_{\dot\alpha}{}^\alpha\psi_\alpha^A\\
 &\hphantom{{}={}\int\de^4x\,\tr\Big({}}{}
 +g_\YM\Bigl(
 \frac{i}{2}\epsilon_{ijk}\phi^i \staracomm{\psi^{\alpha j}}{\psi^k_\alpha}
 +\phi^j \acomm{\bar\psi^{\dot\alpha}_4}{
 \bar\psi_{\dot\alpha j}}_\ast+\text{h.c.}
 \Bigr)\\
 &\phantom{{}={}\int\de^4x\,\tr\Big({}}{}
 -\frac{g^2_\YM}{4}
 \comm{\bar\phi_j}{\phi^j}\comm{\bar\phi_k}{\phi^k}
 +\frac{g^2_\YM}{2}
 \starcomm{\bar\phi_j}{\bar\phi_k}\starcomm{\phi^j}{\phi^k}
 \Big)
  \eqncom
\end{aligned}
\end{equation}
where the spacetime indices $\mu,\nu=0,1,2,3$, spinor indices $\alpha=1,2$, $\dot\alpha=\dot1,\dot2$, flavour indices $i,j,k=1,2,3$ and $A=1,2,3,4$ are implicitly summed over via Einstein's summation convention. The deformation parameters enter via the $\ast$-product which occurs in the $\ast$-deformed commutators and anti-commutators. We have kept only those \cstar-products that do introduce net deformations in the $\gamma_i$-deformation. In the $\beta$-deformation, the antisymmetric product \eqref{eq: antisymmetric product} of the gluino charge vector with any other charge vector also vanishes, yielding $\acomm{\bar\psi^{\dot\alpha}_4}{\bar\psi_{\dot\alpha j}}_\ast|_\beta=\acomm{\bar\psi^{\dot\alpha}_4}{\bar\psi_{\dot\alpha j}}$. 

In addition to the single-trace terms, the deformed theories also contain multi-trace couplings, which are not inherited from the undeformed parent theory. For the deformations with gauge group \SUN, the only additional structures are of double-trace type. For the ones with gauge group \UN, also triple- and quadruple-trace structures may occur. The $\beta$-deformation with gauge group \SUN contains non-vanishing double-trace couplings, which are at the fix-point value such that the theory is conformally invariant \cite{Fokken:2013mza}. They arise when the auxiliary fields, present in the $\mathcal{N}=1$ superspace formulation of the undeformed theory, are integrated out after introducing the deformation, as shown in \cite{Jin:2012np, Fokken:2013aea}.\footnote{The double-trace term follows also directly from the procedure mentioned much earlier in \cite{Freedman:2005cg}.} While these double-trace couplings are absent in the \UN $\beta$-deformation at tree-level, they are induced at loop-level and flow to the \SUN fix point in the infrared, which renders the theory non-conformal. The $\gamma_i$-deformation is not conformally invariant, neither for \UN nor \SUN gauge group. Moreover, conformal invariance cannot be restored by extending the theory with any number of multi-trace couplings.\footnote{See \cite{Fokken:2013aea} for details and possible interpretations in the light of the \AdSCFTc.} The breakdown of conformal invariance originates from running double-trace couplings which have no fix points \cite{Fokken:2013aea}. At one loop, these couplings also affect the planar anomalous dimensions of $L=2$ operators. Hence, starting from two loops, the dilatation operator becomes renormalisation-scheme dependent, as was exemplified in \cite{Fokken:2014soa}. 

As already mentioned in the introduction, the breakdown of conformal invariance does not invalidate our approach at one-loop level. However, it leads to an issue of uniqueness. In the case of the $\cN=1$ supersymmetric $\beta$-deformation, a unique conformally invariant theory exists, which has gauge group $\SUN$. In the case of the $\gamma_i$-deformation, there is no (known) conformally invariant theory. However, a whole family of $\gamma_i$-deformed theories exist that share the single-trace action \eqref{eq: component action single trace} and differ only in the double-trace part of the action. For the sake of definiteness, we focus on the candidate action of the $\gamma_i$-deformation that was proposed in \cite{Frolov:2005dj}, i.e.\ we set the tree-level values of all multi-trace couplings to zero. In the $\beta$-deformation with gauge group $\UN$, we equally set the tree-level values of all multi-trace couplings to zero. This leads to the following double-trace part of the Euclidean action of the $\beta$-deformation
\begin{equation}\label{eq: component action double trace}
\begin{aligned}
S^\beta_{\text{m.t.}} &=\int\de^4x\,\Bigl[-\frac{\colors}{N}
 \frac{g^2_\YM}{2}\tr\bigl(\starcomm{\bar\phi_j}{\bar\phi_k}\bigr)
 \tr\bigl(\starcomm{\phi^j}{\phi^k}\bigr)
\Bigr] \eqncom
\end{aligned}
\end{equation}
where the gauge-group parameter is
\begin{equation}\label{eq: color s}
\colors=\begin{cases}
0\quad\text{ for }\UN\eqncom \\
1\quad\text{ for }\SUN\eqndot
\end{cases}
\end{equation}

The elementary building blocks on which the dilatation operator acts in the planar limit are single-trace operators.\footnote{Interactions that split or join traces are suppressed by $\frac{1}{N}$.} As the deformations do not alter the field content but only the interactions, the single-trace operators are built from the same alphabet as in the undeformed \NfSYMt:
\begin{equation}\label{eq: alphabet}
 \cA =\{ \D^k\phi^i, \D^k\bar\phi^i, \D^k\psi^A_{\alpha}, \D^k\bar\psi^A_{\dot\alpha},\D^k\cF_{\alpha\beta},\D^k\bar\cF_{\dot\alpha\dot\beta} \}
\eqncom
\end{equation}
where the abbreviation $\D^k\bar\psi^2_{\dot\alpha}$ stands for expressions with $k\in \NN_0$ covariant derivatives $\D_{\beta\dot{\beta}}$ acting on $\bar\psi^2_{\dot\alpha}$ and which are totally symmetric in both kinds of spinor indices. These operators can be mapped to cyclic spin-chain states, see \cite{Beisert:2010jr} for a review and appendix \ref{sec:Fields, states and the oscillator picture} for our conventions. The fields in the spin-chain picture can be represented by two sets of bosonic oscillators $\aosc^{\dagger\alpha}$ ($\alpha=1,2$) and $\bosc^{\dagger\alphadot}$ ($\alphadot=1,2$) and one set of fermionic oscillators $\cosc^{\dagger A}$ ($A=1,2,3,4$). They can be characterised by a vector containing the occupation numbers of each oscillator
\begin{equation}\label{eq:occupation_numbers}
A_{i}=(\akindsite[1]{i},\akindsite[2]{i},\bkindsite[1]{i},\bkindsite[2]{i},
\ckindsite[1]{i},\ckindsite[2]{i},\ckindsite[3]{i},\ckindsite[4]{i})\eqncom
\end{equation}
where the lower index $i$ labels the site of the spin chain.

In the 't Hooft limit, the dilatation operator admits a perturbative expansion in the effective planar coupling $g$: 
\begin{equation}\label{eq: lambda def}
D=D_0+g^2 D_2+\cO(g^3)\eqncom \qquad g=\frac{\sqrt\lambda}{4\pi} \eqncom
\end{equation}
where only the classical piece $D_0$ and the one-loop correction $D_2$ are shown. Their actions on single-trace operators, which are represented as spin chains, can be written in terms of site-independent densities $\mathfrak{D}_0$ and $\mathfrak{D}_2$, respectively, as
	\begin{equation}\label{eq: dilatation operator on spin chain}
	\begin{aligned}
	D_{2K}=\sum_{i=1}^L
	\underbrace{\mathds{1}\otimes\dots\otimes\mathds{1}}_{i-1}\otimes\mathfrak{D}_{2K}\otimes\underbrace{\mathds{1}\otimes\dots\otimes\mathds{1}}_{L-i-K}\eqncom
	\end{aligned}
	\end{equation}
where cyclic identification $i+L\sim i$ is understood.

The classical dilatation-operator density $\mathfrak{D}_0$ takes the same form in the deformed theories as in the undeformed one. It yields the classical scaling dimension of the field it acts on. In the spin-chain picture, it takes the diagonal form
\begin{equation}\label{eq: def classical dilatation op in osc language}
(\mathfrak{D}_0)_{A_i}^{A_j}=\Biggl(1+\frac12\sum_{\alpha=1}^{2}\akindsite[\alpha]{i}
+\frac12\sum_{\dot\alpha=\dot 1}^{\dot 2}\bkindsite[\dot{\alpha}]{i}\Biggr)\delta_{A_i}^{A_j}\eqndot
\end{equation}

In contrast to $\mathfrak{D}_0$, the one-loop dilatation-operator density $\mathfrak{D}_2$ depends on the deformation parameters. In the absence of one-loop finite-size effects, which occur at $L=2$ for gauge group \SUN and at $L=1$ for gauge group \UN, it can be expressed in terms of the undeformed density times a phase factor that depends on the order and flavour of the external fields alone \cite{Beisert:2005if}:
\begin{equation}\label{eq: asymptotic dila}
(\mathfrak{D}_2^{L\geq 3})_{A_iA_j}^{A_kA_l}= \e^{\frac{\complexi}{2} (\mathbf{q}_{A_k} \wedge \mathbf{q}_{A_l}- \mathbf{q}_{A_i} \wedge \mathbf{q}_{A_j})}(\mathfrak{D}_2^{\cN=4})_{A_iA_j}^{A_kA_l}\eqndot
\end{equation}
In \cite{Beisert:2003jj}, $\mathfrak{D}_2^{\cN=4}$ was given in terms of all possible hoppings of the oscillators from site $1$ to $2$ and vice versa. Each such hopping combination is weighted by the coefficient\footnote{Note that our coefficient $c_{\textrm{h}}$ has to be divided by 2 to match the conventions of \cite{Beisert:2003jj}.}
\begin{equation}\label{eq: harmonic action coefficient}
c_{\mathrm{h}}(n,n_{12},n_{21})=\begin{cases}2 h(\frac12 n) & \text{if }n_{12}=n_{21}=0 \eqncom\\
2(-1)^{1+n_{12}n_{21}}  
B\left(\frac{n_{12}+n_{21}}{2},1+\frac{n-n_{12}-n_{21}}{2}\right)
& \text{else}\eqncom
\end{cases}
\end{equation}
where $h(m)$ denotes the $m^{\text{th}}$ harmonic number and $B(a,b)$ is the Euler $\beta$-function. This coefficient only depends on the total sum of oscillators at both sites $n$ and the total sum of oscillators that hop from site $i$ to $j$, denoted by $n_{ij}$. In \cite{Fokken:2013mza}, we have given an explicit form of the matrix elements $(\mathfrak{D}_2^{\cN=4})_{A_iA_j}^{A_kA_l}$ in terms of the occupation numbers \eqref{eq:occupation_numbers}, which is also suitable for the present context. 

For short operators, the asymptotic dilatation-operator density \eqref{eq: asymptotic dila} of the deformed theories is altered due to finite-size effects. In the \SUN $\beta$-deformation, they stem from the prewrapping effect \cite{Fokken:2013mza}. It is caused by the double-trace part in the $\SUN$ propagator and the double-trace coupling \eqref{eq: component action double trace}. This effect can be implemented on the level of the dilatation-operator density at one-loop order via a simple prescription \cite{Fokken:2013mza}. It is given as
\begin{equation}\label{eq:D2_beta}
(\mathfrak{D}_2^{L=2})_{A_iA_j}^{A_kA_l}=
\begin{cases}
(\mathfrak{D}_2^{\mathcal{N}=4})_{A_iA_j}^{A_kA_l}&\text{if } 
A_i,A_j,A_k,A_l\in\mathcal{A}_{\text{matter}}\text{ or } A_i,A_j,A_k,A_l\in\bar{\mathcal{A}}_{\text{matter}}\eqncom\\
(\mathfrak{D}_2^{L\geq 3})_{A_iA_j}^{A_kA_l}&\text{else}\eqncom
\end{cases}
\end{equation}
where the sub-alphabets of (anti)-matter fields are defined as
\begin{equation}
\begin{aligned}\label{eq:matter_type}
\cA_{\text{matter}}&=\{ \D^k\phi^1, \D^k\phi^2, \D^k\phi^3, \D^k\psi^1_\alpha, \D^k\psi^2_\alpha, \D^k\psi^3_\alpha \} \eqncom \\
\bar\cA_{\text{matter}}&=\{ \D^k\bar\phi^1, \D^k\bar\phi^2, \D^k\bar\phi^3, \D^k\bar\psi^1_{\dot\alpha}, \D^k\bar\psi^2_{\dot\alpha},\D^k\bar\psi^3_{\dot\alpha} \}
\eqndot
\end{aligned} 
\end{equation}

The one-loop finite-size contributions for the $\beta$- and $\gamma_i$-deformation with gauge group \UN only occur for $L=1$ operators.\footnote{Recall that we are considering the \UN theories with zero tree-level values for all double-trace couplings. The dilatation operator in the presence of multi-trace couplings can also be calculated \cite{FSW}, e.g.\ by an extension of the method of \cite{Wilhelm:2014qua}.} They stem from ordinary wrapping corrections and the corresponding wrapping diagrams are of self-energy type. In appendix D of \cite{Fokken:2013aea}, the one-loop self-energy contributions to the scalar propagators were determined. We present an analogous calculation for self-energy contributions to the fermionic propagators in appendix \ref{app:fermionic self-energies}. The gluon interactions are not affected by the deformations and thus the self-energies for the gluons are undeformed. From the respective contributions, we find the following eigenvalues of $\mathfrak{D}_2$ on $L=1$ operators:
\begin{equation}\label{eq:U1_energies}
\begin{aligned}
\mathrlap{E_{\tr(\D^k\phi^i)}}\hphantom{E_{\tr(\D^k \cF_{\alpha\beta})}}&=
\mathrlap{E_{\tr(\D^k\bar{\phi}^i)}}\hphantom{E_{\tr(\D^k \bar\cF_{\dot\alpha\dot\beta})}}=
4\left(\sin^2\frac{\gamma^+_i}{2}+\sin^2\frac{\gamma^-_i}{2}\right)\eqncom\\
\mathrlap{E_{\tr(\D^k\psi^i_\alpha)}}
\hphantom{E_{\tr(\D^k \cF_{\alpha\beta})}}
&=
\mathrlap{E_{\tr(\D^k\bar\psi^i_{\dot{\alpha}})}}\hphantom{E_{\tr(\D^k \bar\cF_{\dot\alpha\dot\beta})}}=
2\left(\sin^2\frac{\gamma^-_i}{2}+\sin^2\frac{\gamma^+_{i+1}}{2}+\sin^2\frac{\gamma^+_{i+2}}{2}\right)\eqncom\\
\mathrlap{E_{\tr(\D^k \psi^4_\alpha)}}
\hphantom{E_{\tr(\D^k \cF_{\alpha\beta})}}
&=
\mathrlap{E_{\tr(\D^k\bar\psi^4_{\dot{\alpha}})}}\hphantom{E_{\tr(\D^k \bar\cF_{\dot\alpha\dot\beta})}}=
2\sum_{i=1}^3\sin^2\frac{\gamma^-_i}{2}\eqncom\\
E_{\tr(\D^k \cF_{\alpha\beta})}
&=
E_{\tr(\D^k \bar\cF_{\dot\alpha\dot\beta})}=0\eqncom
\end{aligned}
\end{equation}
where cyclic identification $i+3\sim i$ is understood. These results agree with the ones of \cite{Fokken:2013mza} for the $\beta$-deformation in the limit $\gamma_i^+=-\beta$, $\gamma_i^-=0$.

\section{Partition functions via \Polya theory}\label{sec:Partition functions via Polya}
In this section, we summarise the approach of \cite{Spradlin:2004pp} to the one-loop partition function of \NfSYMt via an extension of \Polya theory. We will follow the presentation of \cite{Spradlin:2004pp} and refer the reader there for details. Moreover, we show how this approach can be modified to be applicable in the deformed theories.

In the planar limit, the action of the dilatation operator $D$ on a multi-trace operators is completely determined by the action of $D$ on the operator's single-trace factors. Accordingly, we consider single-trace operators first.

\subsection{The single-trace partition function}
The single-trace partition function is defined in analogy to \eqref{eq: partition function intro D} as trace over all single-trace operators:
\begin{equation}\label{eq: def single-trace partition function}
 Z(x)=\tr_{\RR^4}\left[x^D\right]\eqncom
\end{equation}
where $x=e^{- 1/RT}$ as before. Expanding the dilatation operator in the effective planar coupling $g$ as in \eqref{eq: lambda def}, we obtain the following expansion of \eqref{eq: def single-trace partition function}:
\begin{equation}\label{eq: perturbative expansion of Z}
\begin{aligned}
 Z(x,g) 
&= \tr\left[x^{D_0}\right] + g^2\ln x \tr\left[x^{D_0} D_2\right] + \cO(g^3) \\
&= Z^{(0)}(x) + g^2\ln x\,Z^{(1)}(x)+ \cO(g^3) \eqndot 
\end{aligned}
\end{equation}
Up to the first loop order, the dilatation operator does not change the length $L$ of a single-trace operator. Hence, the respective traces in \eqref{eq: perturbative expansion of Z} can be expressed as sums of traces with fixed $L$. Moreover, the trace in the space of single-trace operator of length $L$ can be re-expressed as trace in the space of spin chains of length $L$ if we include the projector on graded cyclically invariant spin-chain states $\cP$. This projector can be written in terms of the shift operator $T$ as 
\begin{equation}\label{eq: def projector}
 \cP = \frac{1}{L} (1 + T + T^2 + \cdots + T^{L - 1}) \eqndot
\end{equation}

Let us first evaluate the contribution from the free theory
\begin{equation}
 \tr_L[\cP x^{D_0}]\eqndot
\end{equation}
Disregarding the existence of fermions, this problem is equivalent to the enumeration of all necklaces of length $L$ that can be built from a specified alphabet of beads $\cA$. The latter problem can be solved by the enumeration theorem of \Polya \cite{Polya37}. One subtlety arises due to the existence of fermions, which require the necklaces to be graded cyclically invariant.\footnote{If we shift a fermion from the last to the first position in an overall bosonic single-trace operator, we acquire a minus sign.} The final result including this subtlety is
\begin{equation}\label{eq: Polya necklaces for fermions}
 \tr_L[\cP x^{D_0}]= \frac{1}{L}\sum_{k \mid L} \EulerPhi(k) \left[ z(\omega^{k+1} x^k) \right]^{L/k} \eqncom
\end{equation}
where the sum runs over all divisors $k$ of $L$, $\EulerPhi(k)$ is the Euler totient function and 
\begin{equation}\label{eq: single site partition function}
 z(x)= \sum_{A\in \cA} x^{(\mathfrak{D}_0)_A^A} 
\end{equation}
is the single-site partition function. The graded cyclic invariance is incorporated by the formal quantity $\omega$ with the property $\sqrt{\omega}=-1$. It exploits the fact that all fermions have half-integer classical scaling dimensions and it can hence be used in all perturbative expansions \cite{Spradlin:2004pp}. The single-trace partition function at zero coupling is obtained by summing \eqref{eq: Polya necklaces for fermions} over all admissible lengths $L$. Note that this sum has to start at $L=1$ for gauge group $\UN$ and at $L=2$ for gauge group $\SUN$, since $\SUN$ matrices are traceless: 
\begin{equation}\label{eq: free theory single-trace partition function}
 Z^{(0)}(x) = \sum_{L=1+\colors}^\infty \tr_L[\cP x^{D_0}]
= - \colors z(x) - \sum_{k=1}^\infty \frac{\EulerPhi(k)}{k}
\ln[1 - z(\omega^{k+1} x^k)]\eqncom
\end{equation}
with $\colors=0$ for gauge group \UN and $\colors=1$ for gauge group \SUN, as defined in \eqref{eq: color s}.

Now we turn to the first loop order, where we have to evaluate 
\begin{equation}
 \tr_L[\cP x^{D_0}D_2]\eqndot
\end{equation}
Compared to the \NfSYMt case, there is an additional complication due to the fact that the one-loop dilatation-operator density has an explicit $L$-dependence in the deformed theories. It occurs at $L=1$ and $L=2$ and is caused by finite-size effects; cf.\ section \ref{sec:deformed_N4_SYM}. For $L\ge3$, the dilatation-operator density is length-independent and we can use the following result of \cite{Spradlin:2004pp}:\footnote{In the second sum, $(k,L)=1$ denotes that $k$ is relatively prime to $L$.}
\begin{equation}\label{eq: trL result}
\begin{aligned}
 \tr_L[&\cP x^{D_2} {D}_2^{L\ge3}] = \sum_{m \mid L} \EulerPhi(m)\left[ z(\omega^{m+1}x^{m}) \right]^{L/m - 2}
 \ev{ \mathfrak{D}_2^{L\ge3}(\omega^{m+1}x^{m}) } \\
&+ \sum_{\substack{k=0 \\ (k,L) = 1}}^{L-1} \left[\ev{ P\mathfrak{D}_2^{L\ge3}(\omega^{L-k+1}x^{L-k}, \omega^{k+1}x^{k})} - z(\omega^{L+1}x^L)^{-1} \ev{ \mathfrak{D}^{L\ge3}_2(\omega^{L+1}x^L)}\right] \eqncom
 \end{aligned}
\end{equation}
where
\begin{align}
 \label{eq: SV 3.10}
 \ev{ \mathfrak{D}_2^{L\ge3}(x) } &= \sum_{A_1,A_2 \in \cal A} x^{(\mathfrak{D}_0)_{A_1}^{A_1}+(\mathfrak{D}_0)_{A_2}^{A_2}} 
 (\mathfrak{D}_2^{L\ge3})^{ A_1A_2 }_{ A_1A_2 }
  \eqncom \\
 \label{eq: SV 3.11}
 \ev{ P\mathfrak{D}_2^{L\ge3}(w,y) }&=\sum_{A_1,A_2 \in \cal A} (-1)^{F(A_1)F(A_2)} w^{(\mathfrak{D}_0)_{A_1}^{A_1}}y^{(\mathfrak{D}_0)_{A_2}^{A_2}}
  (\mathfrak{D}_2^{L\ge3})^{ A_2A_1 }_{ A_1A_2 } \eqncom 
\end{align}
with $F(A_i)=1$ for fermions and $F(A_i)=0$ for bosons and $\mathfrak{D}_2^{L\ge3}$ given in \eqref{eq: asymptotic dila}. The rich combinatorial structure of this expression originates from the shift operators in \eqref{eq: def projector}, and we refer the reader to \cite{Spradlin:2004pp} for its derivation. At length $L=2$, we have 
\begin{equation}\label{eq: L=2}
 \tr_2[\cP x^{D_2} {D}_2^{L=2}] = \ev{ \mathfrak{D}_2^{L=2}(x)} + \ev{ P\mathfrak{D}_2^{L=2}(x,x)} \eqncom
\end{equation}
and at $L=1$
\begin{equation}\label{eq: L=1}
 \tr_1[\cP x^{D_2} {D}_2^{L=1}] = \sum_{A\in \cal A} x^{(\mathfrak{D}_0)_{A}^{A}}(\mathfrak{D}_2^{L=1})_A^A \eqncom
\end{equation}
with $\mathfrak{D}_2^{L=2}$ and $\mathfrak{D}_2^{L=1}$ given in \eqref{eq:D2_beta} and \eqref{eq:U1_energies}, respectively. The complete one-loop correction to the single-trace partition function is given by the sum of the respective terms over all admissible $L$:
\begin{equation}\label{eq: single-trace one-loop} 
 \begin{aligned}
   Z^{(1)}(x)&= Z^{(1)}_{\text{f.s.c}}(x)+\sum_{L=2}^\infty \tr_L[\cP x^{D_0} D_2^{L\geq 3}]\\
 &= Z^{(1)}_{\text{f.s.c}}(x)+ \sum_{n=1}^\infty \sum_{\substack{k=0 \\ (k,n)=1}}^{n-1} \Bigg[\frac{\ev{\mathfrak{D}_2^{L\ge3}(\omega^{n+1}x^n)}}{1-z(\omega^{n+1}x^n)}  +\delta_{n\neq1} \ev{P\mathfrak{D}_2^{L\geq 3}(\omega^{n-k+1}x^{n-k},\omega^{k+1}x^{k}} \Bigg] \eqncom
 \end{aligned}
\end{equation}
where the finite-size corrections are included in 
\begin{equation}\label{eq:fsc}
 Z^{(1)}_{\text{f.s.c}}(x)= (1-s) \tr_1[\cP x^{D_2} {D}_2^{L=1}] + \tr_2[\cP x^{D_2} {D}_2^{L=2}]-\tr_2[\cP x^{D_2} {D}_2^{L{\geq}3}]
\end{equation}
with the dependence of $\mathfrak{D}_2^{L=2}$ on the gauge group left implicit.

\subsection{The multi-trace partition function}
Multi-trace operators can be build as products of single-trace operators obeying the correct statistics, i.e.\ the Bose-Einstein statistic for bosonic single-trace operators and the Fermi-Dirac statistic for fermionic single-trace operators. For a toy model containing only one single-trace operator $\cO$ with scaling dimension $\Delta$, this leads to the well known  partition function $\left(\frac{1}{1-x^\Delta}\right)$ if $\cO$ is bosonic and $\left(1+x^\Delta\right)$ if $\cO$ is fermionic. The result in the complete theory is the product of these expressions over all single-trace operators. For a perturbative expressing, this can be simplified by the use of $\omega$ introduced below \eqref{eq: single site partition function} to finally arrive at 
\begin{equation}\label{eq: def of multi-trace partition function omega}
  \cZ(x) = \exp \left[ \sum_{n=1}^\infty \frac{Z(\omega^{n+1} x^n)}{n} \right] \eqndot
\end{equation}
Inserting the perturbative expansion of the single-trace partition function \eqref{eq: perturbative expansion of Z} into \eqref{eq: def of multi-trace partition function omega}, we obtain
\begin{equation}
\cZ(x,g)=  \cZ^{(0)}(x)+ g^2 \ln x\, \cZ^{(1)}(x)  + \cO(g^3) \eqncom
\end{equation}
with
\begin{equation}
\begin{aligned}
 \cZ^{(0)}(x) = \exp \left[ \sum_{n=1}^\infty \frac{Z^{(0)}(\omega^{n+1} x^n)}{n} \right] \eqncom \qquad
 \cZ^{(1)}(x)= \cZ^{(0)}(x) \sum_{n=1}^\infty Z^{(1)}(\omega^{n+1} x^n) \eqndot
\end{aligned}
\end{equation}

The final expression for the multi-trace partition function and its one-loop correction can be found by inserting the respective single-trace expressions \eqref{eq: free theory single-trace partition function} and \eqref{eq: single-trace one-loop} into the above equations. After several steps, we obtain
\begin{equation}\label{eq: multi-trace partition function of free nfsymt}
 \cZ^{(0)}(x)
=\exp\left[-\colors\sum_{n=1}^\infty \frac{z(\omega^{n+1}x^n)}{n}\right] \prod_{m=1}^\infty\frac{1}{1-z(\omega^{m+1}x^m)} \eqncom
\end{equation}
and 
\begin{equation}\label{eq: first order of multi-trace partition function}
\begin{aligned}
 \cZ^{(1)}(x)= \cZ^{(0)}(x) \Bigg[\sum_{n=1}^\infty Z_{\text{f.s.c}}(\omega^{n+1} x^n) &+ \sum_{k=1}^{\infty}k\frac{\ev{\mathfrak{D}_2^{L\ge3}(\omega^{k+1}x^{k})}}{1-z(\omega^{k+1}x^k)} \\ &+\sum_{k,m=1}^\infty\ev{P\mathfrak{D}_2^{L\ge3}(\omega^{k+1}x^{k},\omega^{m+1}x^m}\Bigg]\eqndot
 \end{aligned}
\end{equation}

\section{Ingredients}\label{sec:ingredients}
In this section, we compute the ingredients for the extended \Polya-theory method: $z(x)$, $\ev{P\mathfrak{D}_2^{L\geq 3}(w,y)}$, $\ev{\mathfrak{D}^{L\geq 3}_2(x)}$ and $Z^{(1)}_{\text{f.s.c.}}(x)$. We focus on conveying the main ideas and results here and postpone the details to the appendices \ref{app: PD2 calculation}, \ref{app: corrections} and \ref{app: summation identities}.

\subsection{The single-site partition function}\label{sec:single-site partition function}
The single-site partition function defined in \eqref{eq: single site partition function}, as well as all other ingredients, can be derived in the oscillator picture, c.f. appendix \ref{sec:Fields, states and the oscillator picture} for details. In this picture, sums over the full alphabet of the theory \eqref{eq: alphabet} are expressed in terms of sums over all oscillator occupation numbers \eqref{eq:occupation_numbers} as
\begin{equation}\label{eq:alphabet_sum}
\sum_{A_{i}\in \mathcal{A}}= \sum_{\akindsite[1]{i},\akindsite[2]{i}=0}^\infty \, \sum_{\bkindsite[\numberdot1]{i},\bkindsite[\numberdot2]{i}=0}^\infty \,
\sum_{\ckindsite[1]{i},\ckindsite[2]{i},\ckindsite[3]{i},\ckindsite[4]{i}=0}^1 
\delta_{C_{(i)}} \eqndot
\end{equation}
The occurring Kronecker $\delta$ guarantees that only those combinations of oscillators contribute that correspond to one of the fields in the alphabet of the theory \eqref{eq: alphabet}. Its argument is the eigenvalue of the central charge operator defined in \eqref{eq: def central charge operator}.

The single-site partition function only depends on the field content of the theory and thus it is the same for \NfSYMt and the $\beta$- and $\gamma_i$-deformation. Via a direct summation using some of the identities in appendix \ref{app: PD2 calculation}, we find
\begin{equation}\label{eq: single-bead partition function of nfsymt}
  z(x)= \sum_{A\in \mathcal{A}}x^{(\mathfrak{D}_0)_A^A}=\frac{2 \left(3-\sqrt{x}\right) x}{\left(1-\sqrt{x}\right)^3} \eqndot
\end{equation}
This agrees with the result of \cite{Sundborg:1999ue,Spradlin:2004pp,AMMPR03}.

\subsection{\texorpdfstring{The \expectationvalue $\ev{P\mathfrak{D}^{L\geq 3}_2(w,y)}$}{The expectation value <PD2(w,y)>}}\label{sec: PD2}
Next, we compute the permuted expectation value of the asymptotic one-loop dilatation operator. The explicit form of $ \ev{ P\mathfrak{D}^{L\geq 3}_2(w,y) } $ is obtained by inserting \eqref{eq: asymptotic dila}, \eqref{eq:alphabet_sum} and \eqref{eq:PD2_matrix} into \eqref{eq: SV 3.11}. This yields\footnote{Note that in the oscillator picture the fermion number operator takes the form $F(A)=\sum_{e=1}^4c^{e}$. Hence, the factor $(-1)^{F(A_1)F(A_2)}$ in \eqref{eq: SV 3.11} cancels the respective factor from the matrix element in \eqref{eq:PD2_matrix}.}
\begin{equation} 
\begin{aligned}\label{eq: first eqation of PD2}
 \ev{ P\mathfrak{D}^{L\geq 3}_2(w,y) } 
  &=\prod_{i=1}^2\Bigg( \sum_{\akindsite[1]{i},\akindsite[2]{i}=0}^\infty \sum_{\bkindsite[\dot1]{i},\bkindsite[\dot2]{i}=0}^\infty
    \sum_{\ckindsite[1]{i},\ckindsite[2]{i},\ckindsite[3]{i},\ckindsite[4]{i}=0}^1
    \delta_{C_{(i)}} \Bigg)\\
  &\phaneqtimes 
   w^{\frac 12\left(2+\sum_{\alpha=1}^{2}a^\alpha_{(1)}+\sum_{\dot\alpha=\dot 1}^{\dot 2}b^{\dot\alpha}_{(1)}\right)}
   y^{\frac 12\left(2+\sum_{\alpha=1}^{2}a^\alpha_{(2)}+\sum_{\dot\alpha=\dot 1}^{\dot 2}b^{\dot\alpha}_{(2)}\right)}
 \\ &\phaneqtimes
   \e^{-\complexi \sum_{l,m=1}^4 \ckindsite[l]{1}\ckindsite[m]{2}\bq_{\psi^{l}}\wedge\bq_{\psi^{m}}} \\
  &\phaneqtimes \prod_{\alpha=1}^2\bigg(\sum_{\akind[\alpha]=0}^\infty \binom{\akindsite[\alpha]{1}}{\akind[\alpha]}\binom{\akindsite[\alpha]{2}}{\akind[\alpha]}\bigg)
   \prod_{\alphadot=1}^2\bigg(\sum_{\bkind[\alphadot]=0}^\infty \binom{\bkindsite[\alphadot]{1}}{\bkind[\alphadot]}\binom{\bkindsite[\alphadot]{2}}{\bkind[\alphadot]}\bigg)\\
  &\phaneqtimes \prod_{e=1}^4\bigg(\sum_{\ckind[e]=0}^1 \binom{\ckindsite[e]{1}}{{\ckind[e]}}\binom{\ckindsite[e]{2}}{\ckind[e]}(-1)^{\ckind[e]}\bigg)\\
  &\phaneqtimes
   c_{\mathrm{h}}\Bigl[\textstyle \sum_{i=1}^2(
   \sum_{\alpha=1}^{2}a^\alpha_{(i)}+
   \sum_{\dot\alpha=\dot1}^{\dot2}b^{\dot{\alpha}}_{(i)}+
   \sum_{e=1}^{4}c^e_{(i)}),\\ 
   &\phaneq\qquad\quad\,  
   \textstyle \sum_{\alpha=1}^{2}(a^\alpha_{(1)}-a^\alpha)+
   \sum_{\dot\alpha=\dot1}^{\dot2} (b^{\dot\alpha}_{(1)}-b^{\dot\alpha})+ \sum_{e=1}^{4}(c^e_{(1)}-c^e),\\
   &\phaneq\qquad\quad\, 
   \textstyle \sum_{\alpha=1}^{2}(a^\alpha_{(2)}-a^\alpha)+
   \sum_{\dot\alpha=\dot1}^{\dot2} (b^{\dot\alpha}_{(2)}-b^{\dot\alpha})+ \sum_{e=1}^{4}(c^e_{(2)}-c^e)\Bigr]
   \eqncom
\end{aligned}
\end{equation}
where we have also used the antisymmetry of $\mathbf{q}_A\wedge\mathbf{q}_B$. 

Performing the above twelve infinite sums is a decisively complicated task due to their entanglement via the central charge constraint and the coefficient $c_\text{h}$ of the harmonic action. We solve it in three steps which are presented in detail in appendix \ref{app: PD2 calculation}. In a first step, we use that $c_{\mathrm{h}}(n,n_{12},n_{21})$ only depends on the total number of oscillators and the total number of oscillators that change their site. Summation identities of binomial coefficients can hence be employed to cut the number of infinite sums in half. In a second step, we express the coefficient $c_{\mathrm{h}}(n,n_{12},n_{21})$ in terms of the following integral representation
\begin{equation}\label{def:Harmonic_action_integral}
\begin{aligned}
c_{\mathrm{h}}(n,n_{12},n_{21})&= \int_0^1 \measure{t}\Bigl(c^{\text{int}}(n,n_{12},n_{21})-1/t\text{-pole}\Bigr) \eqncom\\
c^{\text{int}}(n,n_{12},n_{21})&=2(-1)^{1+n_{12}n_{21}} t^{\frac12(n_{12}+n_{21})-1} (1-t)^{\frac12(n-n_{12}-n_{21})} \eqncom
\end{aligned}
\end{equation}
where the prescription $-1/t$-pole denotes the subtraction of the $1/t$-pole that occurs when $n_{12}=n_{21}=0$.%
\footnote{A trigonometric version of this integral representation of the harmonic action was proposed in \cite{Zwiebel:2007cpa}.} 
This allows us to reduce the entanglement of the infinite sums by defining
\begin{align}\label{eq:integral_representation_PD2}
 \ev{ P \mathfrak{D}^{L\geq 3}_2(w,y) }_{\text{int}}&=\left.\ev{ P \mathfrak{D}^{L\geq 3}_2(w,y)} \right|_{c_{\mathrm{h}}(n,n_{12},n_{21})\rightarrow c^{\text{int}}(n,n_{12},n_{21})}\eqncom
 \intertext{such that}
  \ev{ P \mathfrak{D}^{L\geq 3}_2(w,y) }&=\int_0^1\measure{t} \left(\ev{ P\mathfrak{D}^{L\geq 3}_2(w,y) }_{\text{int}}-1/t\text{-pole}\right)\eqndot\label{eq:integral_representation_PD2_2}
\end{align}
In a third and last step, we perform the remaining six infinite sums in $\ev{ P \mathfrak{D}^{L\geq 3}_2(w,y) }_{\text{int}}$. Two of them can be eliminated via the central charge constraint. In order to further disentangle the remaining four sums, we reexpress the summand as differential and integral operators acting on simpler expressions such that the sums can be performed and the operators applied afterwards. A minimal example of this procedure is the following:
\begin{equation}
 \sum_{n=0}^\infty (n+1)x^{n}=  \sum_{n=0}^\infty \Diff{x}{} x^{n+1}=  \Diff{x}{} \sum_{n=0}^\infty x^{n+1}=  \Diff{x}{} \frac{x}{1-x}=\frac{1}{(1-x)^2} \eqncom
\end{equation}
where the feasible sum is the geometric series. In the calculation of $\ev{ P \mathfrak{D}^{L\geq 3}_2(w,y) }_{\text{int}}$, however, the final feasible sum has to be evaluated via the generating functions of Legendre polynomials.

The final result is
\begin{equation}
\begin{aligned}\label{eq:Result_PD2}
 \ev{ P\mathfrak{D}^{L\geq 3}_2(w,y) } &=4\Biggl(
\frac{w y(1+w^{1/2})^2(1+y^{1/2})^2}{(1-w^{1/2})^2(1-y^{1/2})^2(w^{1/2}+y^{1/2})^2(1+w^{1/2}y^{1/2})^3}f_1(w,y)\\
&\phantom{{}={}4\Bigl(}+\frac{wy}{(1-w)^2(1-y)^2(1+w^{1/2}y^{1/2})(1-wy)}\sum_{i=1}^3f_2(w,y,\gamma^\pm_i)\\
&\phantom{{}={}4\Bigl(}+f_3(w,y)\ln\left[\frac{1-w}{1-w y}\right]\Biggr)\\
&\phantom{{}={}}+w\leftrightarrow y\eqncom
\end{aligned}
\end{equation}
where
\begin{align}
f_1(w,y)&=2-16w^{1/2}+7w+11w^{1/2}y^{1/2}-16wy^{1/2}+w^{3/2}y^{1/2}+3wy\eqncom\\
f_2(w,y,\gamma^\pm_i)&=
\Bigl(\sin^2\frac{\gamma^+_i}{2}+\sin^2\frac{\gamma^-_i}{2}\Bigr)
\Bigl(12w^{1/2}-4w^{3/2}+4wy^{1/2}-4w^{3/2}y^{1/2}-4w^{3/2}y \nonumber \\*
&\phantom{{}={}}-4w^2y-8w^2y^{3/2}+6w^{1/2}y^{1/2}+6wy-2w^{3/2}y^{3/2}-2w^2y^2\Bigr)
\nonumber \\*
&\phantom{{}={}}+4\sin^2\frac{\gamma^+_i+\gamma^-_i}{2}\Bigl(1+w^{1/2}y^{1/2}-wy-w^{3/2}y^{3/2}\Bigr)\eqncom\\
f_3(w,y)&=-\frac{w(w^{1/2}+3y^{1/2})}{(w^{1/2}+y^{1/2})^3}+\frac{2-6y^{1/2}}{(1-y^{1/2})^3}-\frac{1+3w^{1/2}y^{1/2}}{(1+w^{1/2}y^{1/2})^3}\eqndot
\label{eq:result_functions_last}
\end{align}
The result for the $\beta$-deformation is obtained from \eqref{eq:Result_PD2} for $\gamma_i^-=0$ and $\gamma_i^+=-\beta$. For $\gamma_i^\pm=0$, the second line of \eqref{eq:Result_PD2} drops out and the result of \cite{Spradlin:2004pp} for \NfSYMt is reproduced.\footnote{Note that our conventions for $\mathfrak{D}_2$ and hence also for $ \ev{ P\mathfrak{D}^{L\geq 3}_2(w,y) } $ differ by a factor of $4$ with respect to \cite{Spradlin:2004pp}.}

\subsection{\texorpdfstring{The expectation value $\ev{\mathfrak{D}^{L\geq 3}_2(x)}$}{The expectation value <D2(x)>}}\label{sec: D2}
The expectation value of the one-loop dilatation operator density $\ev{\mathfrak{D}^{L\geq 3}_2(x)}$ can be obtained in analogy to $\ev{P\mathfrak{D}^{L\geq 3}_2(x)}$. In order to apply the above techniques, we define
\begin{equation}
\ev{ \mathfrak{D}^{L\geq 3}_2(w,y)}=\sum_{A_1,A_2 \in \cal A} w^{(\mathfrak{D}_0)_{A_1}^{A_1}}y^{(\mathfrak{D}_0)_{A_2}^{A_2}} (\mathfrak{D}^{L\geq 3}_2 )^{A_1A_2}_{A_1A_2}\eqncom
\end{equation}
which coincides with the original definition in \eqref{eq: SV 3.10}  for $w=y=x$. Note that the matrix element $(\mathfrak{D}^{L\geq 3}_2 )^{A_1A_2}_{A_1A_2}$ is independent of the deformation parameters, as can be seen from \eqref{eq: asymptotic dila}. In particular, $(\mathfrak{D}^{L\geq 3}_2 )^{A_1A_2}_{A_1A_2}=(\mathfrak{D}^{L=2}_2 )^{A_1A_2}_{A_1A_2}=(\mathfrak{D}^{\mathcal{N}=4}_2 )^{A_1A_2}_{A_1A_2}$ such that 
\begin{equation}\label{eq:D2equalities}
\ev{\mathfrak{D}^{L\geq 3}_2(x)}=\ev{\mathfrak{D}^{L=2}_2(x)}=\ev{\mathfrak{D}^{\mathcal{N}=4}_2(x)}\eqndot
\end{equation}
We find\footnote{The matrix element $(\mathfrak{D}^{\mathcal{N}=4}_2 )^{A_1A_2}_{A_1A_2}$ can be obtained from \cite{Fokken:2013mza} by setting $A_{3}=A_{1}$ and $A_{4}=A_{2}$. In addition, we have to shift the summation variables according to
	\begin{equation}
	\begin{aligned}
	\atkind[\alpha]=\akindsite[\alpha]{1}-a^\alpha\eqncom\qquad
	\btkind[\alphadot]=\bkindsite[\alphadot]{1}-b^\alphadot\eqncom\qquad
	\ctkind[e]=\ckindsite[e]{1}-c^e\eqncom
	\end{aligned}
	\end{equation}
which amounts to summing over oscillators that hop from one site to the other instead of oscillators that stay at their original position.}
\begin{equation}\label{eq: D_2 in full theory}
\ev{\mathfrak{D}^{L\geq 3}_2(x)}=4\left(\frac{(1 + \sqrt{x})^2 }{ (1 - \sqrt{x})^6}
\left[ - (1 - 4 \sqrt{x} + x)^2 \ln(1-x) - x (1 - 8 \sqrt{x} + 2 x)\right] \right)\eqndot
\end{equation}
This result agrees with the one of \cite{Spradlin:2004pp}, which was obtained by means of the representation theory of \PSLs44.\footnote{Recall the factor of $4$ difference between our convention for $\mathfrak{D}_2$ and the one of \cite{Spradlin:2004pp}.}

\subsection{\texorpdfstring{The finite-size contributions $Z^{(1)}_{\text{f.s.c.}}(x)$}{The finite size contributions Zfsc}}\label{sec:finite_size_contributions}
Finally, we need to account for the one-loop finite-size effects mentioned in section \ref{sec:deformed_N4_SYM} and calculate $Z^{(1)}_{\text{f.s.c.}}(x)$ defined in \eqref{eq:fsc}. 

For the $\gamma_i$-deformation with gauge group \UN and vanishing tree-level multi-trace couplings, the only finite-size contributions stem from the $L=1$ operators whose anomalous dimensions were given in \eqref{eq:U1_energies}. According to \eqref{eq: L=1}, their contributions to the partition function are 
\begin{equation} 
\begin{aligned}\label{eq: Z_corr2 gamma in full theory}
Z_{\text{f.s.c.}\,\UN}^{(1)}(x,\gamma_i^\pm)&=\sum_{A\in\mathcal{A}}x^{(\mathfrak{D}_0)_{A}^{A}}(\mathfrak{D}_2^{L=1})_A^A
=8\sum_{i=1}^3\left(\sin^2\frac{\gamma_i^+}{2}+\sin^2\frac{\gamma_i^-}{2}\right)\frac{x-x^3+x^{\frac 32}-x^{\frac 52}}{(1-x)^4}\eqncom
\end{aligned} 
\end{equation}
which can be calculated in a similar way as $z(x)$ in subsection \ref{sec:single-site partition function}. The special case of the \UN $\beta$-deformation can be obtained by setting the deformation parameters $\gamma^-_i= 0$ and $\gamma^+_i= -\beta$.

In the $\beta$-deformation with gauge group \SUN, the prewrapping corrections have to be accounted for, i.e.\ instead of the asymptotic density \eqref{eq: asymptotic dila} the finite-size corrected density \eqref{eq:D2_beta} has to be taken at $L=2$. Inserting \eqref{eq: L=2} into \eqref{eq:fsc} and using \eqref{eq:D2equalities}, we find
\begin{equation} 
\begin{aligned}\label{eq: Z_corr in full theory}
 Z_{\text{f.s.c.}\,\SUN}^{(1)}(x,\beta)&=
 \ev{P\mathfrak{D}^{L=2}_2(x,x)} -\ev{P\mathfrak{D}^{L\geq 3}_2(x,x)}
=-6\frac{(x+x^{\frac 32})^2}{(1-x)^4}\left(8\sin^2\frac{\beta}{2}\right)\eqncom
\end{aligned} 
\end{equation}
where the calculation of $\ev{P\mathfrak{D}^{L=2}_2(x,x)} $ is presented in appendix \ref{app: corrections}.

In the $\beta$-deformation, \eqref{eq: Z_corr2 gamma in full theory} and \eqref{eq: Z_corr in full theory} can directly be understood in terms of the anomalous dimensions and characters of the supermultiplets that were identified as affected by the finite-size effects in \cite{Fokken:2013mza}.

\section{Partition function and Hagedorn temperature}\label{sec:Hagedorn temperature}

The one-loop partition function of the $\beta$-deformation with gauge group \UN and \SUN and the $\gamma_i$-deformation with gauge group \UN can be obtained by assembling the ingredients from the previous section. In particular, the final result for the multi-trace partition function is obtained by inserting \eqref{eq: multi-trace partition function of free nfsymt}, \eqref{eq:Result_PD2}, \eqref{eq: D_2 in full theory} and \eqref{eq: Z_corr in full theory} into \eqref{eq: first order of multi-trace partition function}. As it does not allow for significant simplifications, we refrain from showing it.\footnote{The terms in the perturbative expansion of the partition function can be understood via the multiplets of single-trace operators and their one-loop anomalous dimensions. For the $\beta$-deformation, the latter were determined for all operators with classical scaling dimension $\Delta_0\leq 4.5$ in \cite{Fokken:2013mza}.} We have checked the perturbative expansion in $x=\e^{-1/RT}$ for gauge group \UN by modifying the approach of \cite{Mussel:2009uw}. The details of this modification are shown in appendix \ref{app: path integral} and the results of both methods agree. Next, we calculate the Hagedorn temperature.

The Hagedorn temperature $T_\H=T_\H(g=0)$ of free planar \NfSYMt was first calculated in \cite{Sundborg:1999ue}. As $z(x)$ is a monotonically increasing function of $x$, the partition function \eqref{eq: multi-trace partition function of free nfsymt} first diverges if $z(x)=1$. From $z(x_\H)=1$ and \eqref{eq: single-bead partition function of nfsymt}, we find
\begin{equation}\label{eq: zero-loop Hagedorn temperature}
 T_\H= \frac{1}{\ln(7+4\sqrt{3})}\frac 1R \eqndot
\end{equation}
This temperature is independent of the deformation parameters, as the free $\beta$- and $\gamma_i$-deformation coincide with the free \NfSYMt.
Remarkably, the one-loop correction to the Hagedorn temperature is also independent of the deformation parameters --- as we will show in the following.

For \NfSYMt, the one-loop correction to the Hagedorn temperature was calculated in \cite{Spradlin:2004pp}. At this temperature, the partition function has a simple pole:
\begin{equation}
\cZ(x) \sim \frac{C}{x_\H-x}\eqncom 
\end{equation}
with a constant $C$. Expanding it around the pole, one finds
\begin{equation}\label{eq: hagedorn temperature expansion}
 \frac{C}{x_\H+\delta x_\H-x}=\frac{C}{x_\H-x}\left[1-\frac{\delta x_\H}{x_\H-x}+\dots\right] \eqndot
\end{equation}
For the deformed theories, we follow the reasoning of \cite{Spradlin:2004pp} and compare the expansion \eqref{eq: hagedorn temperature expansion} to the multi-trace partition function \eqref{eq: first order of multi-trace partition function}. As in the undeformed case, we find that only the $k=1$ term in the second sum of \eqref{eq: first order of multi-trace partition function} contributes to the double pole which shifts the Hagedorn temperature at one loop. In principle, the first and third sum could also develop divergences when evaluated. In the \su{2}, \so{6} and \su{2|3} subsectors, this does, however, not occur; the first and third sum can be evaluated in a closed form and their contribution is finite for finite temperatures. Numerical studies suggest that their contribution in the full theory remains finite at $T_\H$ as well.

As $\ev{\mathfrak{D}^{L\geq 3}_2(x)}$ and $z(x)$ are undeformed in the $\beta$- and \gidef, so is the one-loop correction to the Hagedorn temperature. From the residue of the aforementioned $k=1$ term, we obtain
\begin{equation}
 \delta x_\H= - \lim_{x\rightarrow x_\H} \left[g^2 (x_\H-x) \ln x\frac{\ev{\mathfrak{D}^{L\geq 3}_2(x)}}{1-z(x)}\right]=-\frac{2}{3}g^2x_\H  \ln x_\H\ev{\mathfrak{D}^{L\geq 3}_2(x_\H)} \eqndot
\end{equation}
Inserting \eqref{eq: zero-loop Hagedorn temperature} into \eqref{eq: D_2 in full theory} yields $\ev{\mathfrak{D}^{L\geq 3}_2(x_\H)}=3$. 
Thus,
\begin{equation}
 \frac{\delta x_\H}{x_\H}=-2 g^2\ln x_\H  
\end{equation}
and 
\begin{equation}
 \frac{\delta T_\H}{T_\H}=-\frac{1}{\ln x_\H}\frac{\delta x_\H}{x_\H}= 2 g^2\eqndot
\end{equation}
Hence, the one-loop Hagedorn temperature of $\beta$- and $\gamma_i$-deformed \NfSYMt is given by
\begin{equation}
 T_\H(g)= T_\H\left(1+2 g^2+ \dots \right) \eqncom\qquad g^2=\frac{\lambda}{(4\pi)^2}=\frac{Ng_\YM^2}{(4\pi)^2}\eqncom
\end{equation}
which is identical to the one-loop Hagedorn temperature of \NfSYMt as computed in \cite{Spradlin:2004pp}.

\section{Conclusion and outlook}\label{sec:results}
In this paper, we have calculated the one-loop partition function of the $\beta$- and $\gamma_i$-de\-for\-ma\-tion of \NfSYMt on \RxSt.

For the computation of the partition function, we have used an extension of the generalised \Polya-theory method of \cite{Spradlin:2004pp}, which is based on the dilatation operator of the theory on $\RR^{4}$. Separating off the combinatorial problem, the thermal one-loop partition function of \NfSYMt can be given in terms of three physical ingredients alone: the single-site partition function $z(x)$ and the generalised expectation values $\ev{\mathfrak{D}^{L\geq 3}_2(x)}$ and $\ev{P\mathfrak{D}^{L\geq 3}_2(w,y)}$. In the $\beta$- and $\gamma_i$-deformed theories, the thermal one-loop partition function in addition depends on the finite-size contributions which arise from the length-dependent contributions of the respective dilatation operator. We have calculated all four ingredients in the deformed theories in a purely combinatorial approach. In the limit of vanishing deformation parameters, we reproduce the results of \cite{Spradlin:2004pp} and also provide an analytic derivation for $\ev{P\mathfrak{D}^{L\geq 3}_2(w,y)}$.

We have checked our result for the one-loop partition function in the case of gauge group \UN by modifying the direct path-integral\,/\,Feynman-diagram calculation \cite{Mussel:2009uw} on \RxSt to include the $\beta$- and \gidef. The results of both methods agree.

The space-time manifold \RxSt, with small radius $R$, enables a meaningful treatment of thermodynamic phenomena in gauge theories via perturbation theory. The maximally supersymmetric \NfSYMt as well its $\beta$- and $\gamma_i$-deformations exhibit a Hagedorn transition that was identified with the confinement-deconfinement phase transition in the free theory. While the temperature of this phase transition is trivially independent of the deformations in the free theory, this independence is not guaranteed at the $\cO(\lambda)$ correction, where the dilatation operator becomes deformation-parameter dependent. Remarkably, we have found that the $\cO(\lambda)$ correction to the Hagedorn temperature is independent of the deformation parameters as well.

In \cite{Gursoy06}, the thermodynamic properties of a certain one-parameter deformed background in string theory were investigated. The Hagedorn temperature was found to be undeformed although the partition function depends non-trivially on the deformation parameter. In subsequent research, it was tried to repeat this analysis for the string-theory dual to the $\beta$-deformation, but the analysis was only successful in sectors that do not allow for non-trivial tests \cite{HMP07}. Motivated by our results and the findings on the string theory side, it is tempting to propose that the Hagedorn temperature of the $\beta$-deformation is independent of the deformation parameters at all values of $\lambda$. Further investigations are clearly required.

It would also be interesting to compute the partition function and the Hagedorn temperature in \NfSYMt to two-loop order. In \cite{GRNS05}, it was shown how to generalise the method of \cite{Spradlin:2004pp} to two loops in the \su{2} sector and a generalisation to two or even more loops in the full theory seems feasible. Once the complete two-loop dilatation operator of \NfSYMt becomes available, computing the two-loop partition function of \NfSYMt reduces to a combinatorial exercise. Moreover, it would be interesting to determine the order of the phase transition in \NfSYMt like it was done in \cite{AMMPR05} for pure Yang-Mills theory on \RxSt.

In \cite{Spradlin:2004pp}, it was remarked that the \Polya-theory approach to partition functions makes no apparent use of the integrability observed in \NfSYMt. Naively, this is not surprising: the Bethe ansatz is a powerful tool for finding eigenvalues of $D$ and hence of $x^D$, i.e.\ roots of the characteristic polynomial, while the partition function is the trace of $x^D$, which is determined by the second to highest term of the characteristic polynomial alone. However, the whole spectrum of \NfSYMt is --- in principle --- known via integrability and by knowing the spectrum the partition function can be constructed immediately. It would thus be very interesting to develop a formalism that directly employs the methods of integrability to derive the thermal partition function of \NfSYMt. This might even lead to a closed all-loop expression.


\section*{Acknowledgements}
It is a pleasure to thank our adviser Matthias Staudacher and Yumi Ko for numerous discussions throughout the project and in particular Matthias Staudacher for suggesting it. 
We are very grateful to Christoph Sieg for various discussions, fruitful collaboration on related projects and for sharing his conventions for initial Feynman diagram calculations.
We thank Ofer Aharony and Ran Yacobi for a very helpful discussion on the modifications of the path-integral approach on $\text{S}^1\times\text{S}^3$ to include the deformed theories with gauge group \UN.
Moreover, we thank
Burkhard Eden,
Vladimir Mitev,
Elli Pomoni,
Radu Roiban 
and
Marcus Spradlin
for discussions and communications on various aspects of this work. 
We thank the Israel Institute for Advanced Studies, Jerusalem, for warm hospitality during a visit in 2012, where part of the research presented in this paper was done.
Our work was supported by the SFB 647 \emph{``Raum-Zeit-Materie. Analytische und Geometrische Strukturen''} and the Marie Curie network GATIS (\texttt{\href{http://gatis.desy.eu}{gatis.desy.eu}}) of the European Union’s Seventh Framework Programme FP7/2007-2013/ under REA Grant Agreement No 317089. Wir danken der Studienstiftung des deutschen Volkes für Promotionsförderstipendien.

\appendix
\numberwithin{equation}{section}
\section{\texorpdfstring{The one-loop anomalous dimensions of fermionic $L=1$ operators}{The one-loop anomalous dimensions of fermionic L=1 operators}}\label{app:fermionic self-energies}
\begin{fmffile}{appendix}
\feynmpdefinitionsdiags
In this appendix, we calculate the one-loop self-energies of the fermions, from which we obtain the anomalous dimensions of the fermionic $L=1$ operators in the deformed theories with gauge group \UN. 

Using the Feynman rules of \cite{Fokken:2013aea} and Fermi-Feynman gauge, we arrive at the following divergent one-loop contributions to the fermionic propagator
\begin{equation}
\begin{aligned}
\Kop\Bigl[\,\,\aswfonelabelf{dashes_sarrow}{plain_srarrow,left=1}{dashes_sarrow}{$\scriptstyle \alpha A a$}{$\scriptstyle \dot{\beta} B b$}\,\,\Bigr]\,&=
-\frac{g^2}{\varepsilon}\delta_A^B(1-\delta_A^4)
\Big[2\tr(\T^a\T^b)-\frac 1N\tr(\T^a)\tr(\T^b)\smash{\sum_{\substack{k=1\\k\neq A}}^3}\cos\gamma^+_k\Big]p_{\dot{\beta}}{}^\alpha\eqncom\\
\Kop\Bigl[\,\,\aswfonelabelf{dashes_sarrow}{plain_sarrow,left=1}{dashes_sarrow}{$\scriptstyle \alpha A a$}{$\scriptstyle \dot{\beta}B b$}\,\,\Bigr]\,&=-\frac{g^2}{\varepsilon}\delta_A^B
\Big[(1+2\delta_{A}^4)\tr(\T^a\T^b)\\
&\phantom{=-\frac{g^2}{\varepsilon}\delta_A^B
	\Big[}-\frac 1N\tr(\T^a)\tr(\T^b)\smash{\sum_{k=1}^3}(\delta_{A}^k+\delta_{A}^4)\cos\gamma^-_k\Big]p_{\dot{\beta}}{}^\alpha\eqncom\\
\Kop\Bigl[\,\,\aswfonelabelf{dashes_sarrow}{photon,left=1}{dashes_srarrow}{$\scriptstyle \alpha A a$}{$\scriptstyle \dot{\beta} B b$}\,\, \Bigr]\,&=-\frac{g^2}{\varepsilon}\delta_A^B
\Big[\tr(\T^a\T^b)-\frac 1N\tr(\T^a)\tr(\T^b)\Big]p_{\dot{\beta}}{}^\alpha\eqncom
\end{aligned}
\end{equation}
where the operator $\Kop$ extracts the UV divergence of the respective diagram and $g$ was defined in \eqref{eq: dilatation operator on spin chain}. Taking the sum of the three contributions yields the counterterm for the \U1 component of the fermionic fields\footnote{Note that $\tr(\T^a\T^b)=\delta^{ab}$, $\tr(\T^a)=\sqrt{N}\delta^{a0}$ and the inverse of the propagator $-\frac{p_\alpha{}^{\dot{\beta}}}{p^2}$ is $\frac{p_{\dot{\beta}}{}^\alpha}{2}$.} 
\begin{equation}
\begin{aligned}
\delta \mathcal{Z}^{(1)}_{\psi^A,\U1}&=-\frac{g^2}{\varepsilon}\left(3
-\sum_{k=1}^3\left(\cos\gamma^+_k+(\delta_A^k+\delta_A^4)(\cos\gamma^-_k-\cos\gamma^+_k)\right)
\right)\eqndot
\end{aligned}
\end{equation} 
The trace of an $L=1$ operator projects to the \U1 component and the covariant derivatives reduce to the ordinary ones in this case. Moreover, all diagrams contributing to the anomalous dimension of such operators are of one-particle-reducible type. Hence, the operator renormalisation constant is simply given by
\begin{equation}
\begin{aligned}
\mathcal{Z}_{\tr(\D^k\psi^A)}=\mathcal{Z}^{-\frac 12}_{\psi^A,\U1}\eqndot
\end{aligned}
\end{equation} 
Accordingly, the required one-loop anomalous dimensions are given by
\begin{equation}
\begin{aligned}
\gamma_{\tr(\D^k\psi^A)}^{(1)}&=
\lim_{\epsilon\rightarrow 0}\left(g\epsilon\frac{\de}{\de g}\ln \mathcal{Z}^{-\frac 12}_{\psi^A,\U1}\right)^{(1)}\\
&=2 \sum_{k=1}^3\left[\sin^2\frac{\gamma_k^+}{2}+(\delta_A^k+\delta_A^4)
\left(\sin^2\frac{\gamma^-_k}{2}-\sin^2\frac{\gamma^+_k}{2}\right)\right]\eqndot
\end{aligned}
\end{equation}

\end{fmffile}

\section{The oscillator picture}\label{sec:Fields, states and the oscillator picture}
In this appendix, we present our conventions for the oscillator picture \cite{Gunaydin82,Gunaydin:1998sw,Beisert:2003jj} and the action of the undeformed one-loop dilatation-operator density therein.

The fields from the alphabet \eqref{eq: alphabet} can be represented via two sets of bosonic oscillators $\aosc^{\dagger\alpha}$ ($\alpha=1,2$) and $\bosc^{\dagger\alphadot}$ ($\alphadot=1,2$) and one set of fermionic oscillators $\cosc^{\dagger A}$ ($A=1,2,3,4$). These oscillators obey the usual (anti-)commutation relations:
\begin{equation}\label{eq: (anti)commutation relations}
[\aosc_{\alpha},\aosc^{\dagger\beta}]=\delta_\alpha^\beta \eqncom \qquad
[\bosc_{\dot\alpha},\bosc^{\dagger\dot\beta}]=\delta_{\dot\alpha}^{\dot\beta} \eqncom \qquad
\{\cosc_A,\cosc^{\dagger B}\}=\delta_A^B\eqncom  
\end{equation}
with all other (anti-)commutators vanishing. In terms of the oscillators, the fields are 
\begin{equation}\label{eq: fields}
\begin{aligned}
\cder^k \cfstrength^{\phantom{ABC}} &\mathrel{\widehat{=}} 
  (\aoscdag)^{k+2} 
  (\boscdag)^{k\phantom{+0}}
  \vac \eqncom \\
\cder^k \ferm^{A\phantom{BC}} &\mathrel{\widehat{=}}     
  (\aoscdag)^{k+1} 
  (\boscdag)^{k\phantom{+0}}
  \cosc^{\dagger A}
  \vac \eqncom \\
\cder^k \sufphi^{AB\phantom{C}} &\mathrel{\widehat{=}}     
  (\aoscdag)^{k\phantom{+0}} 
  (\boscdag)^{k\phantom{+0}} 
  \cosc^{\dagger A} \cosc^{\dagger B}
  \vac \eqncom \\
\cder^k \antiferm^{ABC} &\mathrel{\widehat{=}} 
  (\aoscdag)^{k\phantom{+0}} 
  (\boscdag)^{k+1} 
  \cosc^{\dagger A} \cosc^{\dagger B} \cosc^{\dagger C} 
  \vac \eqncom \\
\cder^k \cantifstrength^{\phantom{ABC}} &\mathrel{\widehat{=}}   
  (\aoscdag)^{k\phantom{+0}}
  (\boscdag)^{k+2} 
  \cosc^{\dagger 1} \cosc^{\dagger 2} \cosc^{\dagger 3} \cosc^{\dagger 4}
  \vac \eqncom
\end{aligned}
\end{equation}
where $\antiferm_{D}=\frac{1}{3!}\varepsilon_{ABCD}\antiferm^{ABC}$ and we have suppressed all spinor indices. All physical fields fulfil the central charge constraint, i.e.\ the action of the following operator on them vanishes:
\begin{equation}\label{eq: def central charge operator}
\centralchargeopdensity=\sum_{\alpha=1}^2\aosc^{\dagger\alpha} \aosc_\alpha-\sum_{\alphadot=\numberdot{1}}^{\numberdot{2}}\bosc^{\dagger\alphadot} \bosc_\alphadot+\sum_{A=1}^4 \cosc^{\dagger A}\cosc_A-2\eqndot
\end{equation} 
A generic $n$-site state is given by $n$ families of oscillators $(\aosc^{\dagger \alpha}_{(i)},\bosc^{\dagger\alphadot}_{(i)},\cosc^{\dagger A}_{(i)})$ individually satisfying \eqref{eq: def central charge operator}. It can be represented in terms of its oscillator occupation numbers
\begin{equation}\label{eq:oscillator_occupation_numbers}
A_{i}=(\akindsite[1]{i},\akindsite[2]{i},\bkindsite[1]{i},\bkindsite[2]{i},
\ckindsite[1]{i},\ckindsite[2]{i},\ckindsite[3]{i},\ckindsite[4]{i})\eqncom
\end{equation}
where the index $i=1,\dots,n$ specifies the site on which the corresponding oscillators act. 

In \cite{Beisert:2003jj}, the harmonic action is given as a weighted sum over all reorderings of the oscillators at two neighbouring sites with weight \eqref{eq: harmonic action coefficient}. An explicit expression in terms of the occupation numbers \eqref{eq:oscillator_occupation_numbers} can be found in \cite{Fokken:2013mza}. Restricting the expression (C.4) in \cite{Fokken:2013mza} to the permuted diagonal element, i.e.\ $A_{3}=A_{2}$ and $A_{4}=A_{1}$, we find
\begin{equation}
\begin{aligned}\label{eq:PD2_matrix}
(\mathfrak{D}_2^{\mathcal{N}=4} )^{A_2A_1}_{A_1A_2}&= 
\prod_{\alpha=1}^2\bigg(\sum_{\akind[\alpha]=0}^\infty 
\binom{\akindsite[\alpha]{1}}{\akind[\alpha]}
\binom{\akindsite[\alpha]{2}}{\akindsite[\alpha]{2}-\akind[\alpha]}\bigg)
 \prod_{\alphadot=1}^2\bigg(\sum_{\bkind[\alphadot]=0}^\infty 
 \binom{\bkindsite[\alphadot]{1}}{\bkind[\alphadot]}
 \binom{\bkindsite[\alphadot]{2}}{\bkindsite[\alphadot]{2}-\bkind[\alphadot]}\bigg)\\
&\phaneqtimes \prod_{e=1}^4\bigg(\sum_{\ckind[e]=0}^1 
\binom{\ckindsite[e]{1}}{{\ckind[e]}}
\binom{\ckindsite[e]{2}}{\ckindsite[e]{2}-\ckind[e]}\bigg)
(-1)^{\sum_{e=1}^4\sum_{l=1}^4c^e_{(1)}c^l_{(2)}+\ckind[e]}\\
&\phaneqtimes
 c_{\mathrm{h}}\Bigl[\textstyle \sum_{i=1}^2(
 \sum_{\alpha=1}^{2}a^\alpha_{(i)}+
 \sum_{\dot\alpha=\dot1}^{\dot2}b^{\dot{\alpha}}_{(i)}+
 \sum_{e=1}^{4}c^e_{(i)}),\\ 
 &\phaneq\qquad\quad\,  
\textstyle \sum_{\alpha=1}^{2}(a^\alpha_{(1)}-a^\alpha)+
 \sum_{\dot\alpha=\dot1}^{\dot2} (b^{\dot\alpha}_{(1)}-b^{\dot\alpha})+ \sum_{e=1}^{4}(c^e_{(1)}-c^e),\\
 &\phaneq\qquad\quad\, 
 \textstyle \sum_{\alpha=1}^{2}(a^\alpha_{(2)}-a^\alpha)+
  \sum_{\dot\alpha=\dot1}^{\dot2} (b^{\dot\alpha}_{(2)}-b^{\dot\alpha})+ \sum_{e=1}^{4}(c^e_{(2)}-c^e)\Bigr]
 \eqncom
\end{aligned}
\end{equation}
where $a^\alpha$, $b^\alpha$ and $c^e$ are the numbers of oscillators that stay at their initial sites and $c_{\mathrm{h}}$ is given in \eqref{eq: harmonic action coefficient}.

\section{The calculation of \texorpdfstring{$\ev{P\mathfrak{D}^{L\geq3}_2(w,y)}$}{<PD2(L>=3)(w,y)>}}\label{app: PD2 calculation}
In section \ref{sec:ingredients}, we have sketched the computation of $\ev{P\mathfrak{D}^{L\geq3}_2(w,y)}$. In this appendix, we present the details of this computation in three steps, starting from \eqref{eq: first eqation of PD2}. 

In the first step, we use that the harmonic action $c_{\text{h}}$ is insensitive to the kind of oscillator that is hopping. We rewrite the occurring bosonic summations in terms of the variables
\begin{equation}
\begin{aligned}
\akindsite{i}=\sum_{\alpha=1}^2\akindsite[\alpha]{i}\eqncom\quad
a=\sum_{\alpha=1}^2\akind^{\alpha}\eqncom\quad
\bkindsite{i}&=\sum_{\dot\alpha=\dot1}^{\dot2}\bkindsite[\dot\alpha]{i}\eqncom\quad
b=\sum_{\dot\alpha=\dot1}^{\dot2}\bkind^{\dot\alpha}\eqncom
\end{aligned}
\end{equation}
for the sites $i=1,2$. For each pair $(a_{\bullet}^1,a_{\bullet}^2)\in\{(\akindsite[1]{1},\akindsite[2]{1}),
(\akindsite[1]{2},\akindsite[2]{2}),(\akind[1],\akind[2])\}$, we use the summation idenity
\begin{equation}\label{eq:sum_order}
\sum_{a_{\bullet}^1,a_{\bullet}^2 = 0}^\infty f(a_{\bullet}^1,a_{\bullet}^2) = \sum_{a_{\bullet}= 0}^\infty \sum_{\tilde a_{\bullet}=0}^{a_{\bullet}} f(\tilde a_{\bullet},a_{\bullet}-\tilde a_{\bullet})\eqncom
\end{equation}
which is valid for any function $f$, to express all occurences of $\akind_\bullet^\alpha$ in terms of $a_\bullet$ and $\tilde a_\bullet$. In the resulting expressions, the sums over $\tilde{\akind}_{(i)}$ can be performed via the identity
\begin{equation}
\sum_{\tilde{\akind}_{(i)}=0}^{\akindsite{i}} \binom{\tilde{\akind}_{(i)}}{\tilde{\akind}} \binom{\akindsite{i}-\tilde{\akind}_{(i)}}{\akind-\tilde\akind}= \binom{\akindsite{i}+1}{\akind+1}\eqncom
\end{equation}
and the remaining sum over $\tilde\akind$ simply yields a factor of $(\akind+1)$. This procedure allows us to directly perform three of the six original sums over $\aosc$-type oscillators. The sums involving $\bosc$-type oscillators can be treated analogously. While the coefficient $c_{\text{h}}$ is independent of the kind of $\cosc$-type oscillator that is hopping, the phase factor which incorporates the deformation depends on it. We define
\begin{equation} 
\begin{aligned}\label{eq: def G(c1,c2,c)}
 G^{\gamma_i}(\ckindsite1,\ckindsite2,\ckind)&= \prod_{e=1}^4\bigg(\sum_{\ckindsite[e]{1},\ckindsite[e]{2},\ckind[e]=0}^1 \binom{\ckindsite[e]{1}}{{\ckind[e]}}\binom{\ckindsite[e]{2}}{\ckind[e]}(-1)^{\ckind[e]}\bigg) 
 \e^{-\complexi \sum_{l,m=1}^4 \ckindsite[l]{1}\ckindsite[m]{2}\bq_{\psi^{l}}\times\bq_{\psi^{m}}} \\
 &\qquad \times \delta_{\left(\ckindsite{1}-\sum_{e=1}^4\ckindsite[e]{1}\right)}
 \delta_{\left(\ckindsite{2}-\sum_{e=1}^4\ckindsite[e]{2}\right)}
 \delta_{\left(\ckind-\sum_{e=1}^4\ckind[e]\right)}\eqndot
\end{aligned}
\end{equation}
With the above simplifications, \eqref{eq: first eqation of PD2} turns into
\begin{equation}\label{eq: PD2 pre dual way}
\begin{aligned}
 \ev{P\mathfrak{D}^{L\geq3}_2(w,y)}&= \sum_{\akindsite{1},\akindsite{2},\akind=0}^\infty \sum_{\bkindsite{1},\bkindsite{2},\bkind=0}^\infty \sum_{\ckindsite{1},\ckindsite{2},\ckind=0}^4
 \delta_{(\akindsite1-\bkindsite1+\ckindsite1-2)}\delta_{(\akindsite2-\bkindsite2+\ckindsite2-2)}\\
 &\phaneqtimes w^{\frac12\left(2+\akindsite{1}+\bkindsite{1}\right)} 
 y^{\frac12\left(2+\akindsite{2}+\bkindsite{2}\right)}\, G^{\gamma_i}(\ckindsite1, \ckindsite2, \ckind) \\
&\phaneqtimes (\akind+1)\binom{\akindsite{1}+1}{\akind+1}\binom{\akindsite{2}+1}{\akind+1} (\bkind+1)\binom{\bkindsite{1}+1}{\bkind+1}\binom{\bkindsite{2}+1}{\bkind+1} \\
&\phaneqtimes
 c_{\mathrm{h}}\Bigl[\textstyle \sum_{i=1}^2(a_{(i)}+b_{(i)}+c_{(i)}),\\ 
 &\phaneq\qquad\quad\, (a_{(1)}-a)+(b_{(1)}-b)+(c_{(1)}-c),\\
 &\phaneq\qquad\quad\, (a_{(2)}-a)+(b_{(2)}-b)+(c_{(2)}-c)\Bigr]\eqndot
\end{aligned}
\end{equation}
We can further use the Kronecker $\delta$'s from the central charge constraint to eliminate two of the remaining sums, say those over $b_{(1)}$ and $b_{(2)}$.

In the second step, we employ the integral representation of the harmonic action \eqref{def:Harmonic_action_integral} to replace \eqref{eq: PD2 pre dual way} by the respective integrand defined in \eqref{eq:integral_representation_PD2}:
\begin{equation}\label{eq: PD2 after delta summation}
\begin{aligned}
 \ev{ P \mathfrak{D}_2(w,y) }_{\text{int}}&= -2 \sum_{\ckindsite{1},\ckindsite{2},\ckind=0}^4 \sum_{\akind,\bkind=0}^\infty  \sum_{\akindsite{1}=\maxset{0,2-\ckindsite{1}}}^\infty \sum_{\akindsite{2}=\maxset{0,2-\ckindsite{2}}}^\infty \\
 &\qquad \times G^{\gamma_i}(\ckindsite{1}, \ckindsite{2}, \ckind) w^{\akindsite{1} + \frac12\ckindsite{1}} y^{\akindsite{2} + \frac12\ckindsite{2}} \\
 &\qquad \times  (\akind+1)\binom{\akindsite{1}+1}{\akind+1} \binom{\akindsite{2}+1}{\akind+1}\\
 &\qquad \times  (\bkind+1)\binom{\akindsite{1}+\ckindsite{1}-1}{\bkind+1} \binom{\akindsite{2}+\ckindsite{2}-1}{\bkind+1} \\
 &\qquad \times  t^{\akindsite{1}+\ckindsite{1}+\akindsite{2}+\ckindsite{2}-3} \left(\frac{t-1}{t}\right)^{\akind+\bkind+\ckind}\eqndot
 \end{aligned}
\end{equation}

This brings us to the third step. The combinatorial coefficients occurring in \eqref{eq: PD2 after delta summation} can be rewritten in terms of differential and integral operators acting on the thermal weights $w$ and $y$:
\begin{equation}
\begin{aligned}
 \ev{ P \mathfrak{D}_2(w,y) }_{\text{int}}&= -2 \sum_{\ckindsite{1},\ckindsite{2},\ckind=0}^4  \sum_{\akind,\bkind=0}^\infty  \sum_{\akindsite{1}=\maxset{0,2-\ckindsite{1}}}^\infty \sum_{\akindsite{2}=\maxset{0,2-\ckindsite{2}}}^\infty \\
&\qquad \times  G^{\gamma_i}(\ckindsite{1}, \ckindsite{2}, \ckind) \frac{1}{\akind!(\akind+1)!}\frac{1}{\bkind!(\bkind+1)!} \\
&\qquad \times  \left( w^{-\frac12\ckindsite{1}} \IntOp{w}{0}{w} w^{\frac12\ckindsite{1}-1} \right)  \left( w^{-\frac12\ckindsite{1}} w^{\bkind+2}\Diff{w}{\bkind+2} w^{\frac12\ckindsite{1}} \right) \\
&\qquad \times \left(w^{\frac12\ckindsite{1}-1}w^{\akind+1}\Diff{w}{\akind+1} w^{1-\frac12\ckindsite{1}} \right) w^{\akindsite{1} + \frac12\ckindsite{1}} \\
&\qquad \times \left( y^{-\frac12\ckindsite{2}} \IntOp{y}{0}{y} y^{\frac12\ckindsite{2}-1} \right) \left( y^{-\frac12\ckindsite{2}} y^{\bkind+2}\Diff{y}{\bkind+2} y^{\frac12\ckindsite{2}} \right) \\
&\qquad \times \left(y^{\frac12\ckindsite{2}-1}y^{\akind+1}\Diff{y}{\akind+1} y^{1-\frac12\ckindsite{2}} \right) y^{\akindsite{2} +\frac12\ckindsite{2}} \\
&\qquad \times  t^{\akindsite{1}+\ckindsite{1}+\akindsite{2}+\ckindsite{2}-3} \left(\frac{t-1}{t}\right)^{\akind+\bkind+\ckind}\eqncom
\end{aligned}
\end{equation}
where both differential and integral operators act on everything on their right.\footnote{Note that, in a slight abuse of notation, we have labelled the integration variable with the same symbol as the upper integration boundary.} Since those operators do not explicitly depend on $\akindsite{1}$ and $\akindsite{2}$, we can now perform the sums over these two variables using the well known result for geometric series:
\begin{equation}\label{eq:PD2_diff_Operators}
\begin{aligned} 
 \ev{ P \mathfrak{D}_2(w,y) }_{\text{int}}&= -2 \sum_{\ckindsite{1},\ckindsite{2},\ckind=0}^4  \sum_{\akind,\bkind=0}^\infty    G^{\gamma_i}(\ckindsite{1}, \ckindsite{2}, \ckind) \frac{1}{\akind!(\akind+1)!}\frac{1}{\bkind!(\bkind+1)!} \\
&\qquad \times  \left( w^{-\frac12\ckindsite{1}} \IntOp{w}{0}{w} w^{\frac12\ckindsite{1}-1} \right)  \left( w^{-\frac12\ckindsite{1}} w^{\bkind+2}\Diff{w}{\bkind+2} w^{\frac12\ckindsite{1}} \right) \\
&\qquad \times \left( w^{\frac12\ckindsite{1}-1}w^{\akind+1}\Diff{w}{\akind+1} w^{1-\frac12\ckindsite{1}} \right) w^{\frac12\ckindsite{1}} \frac{(w t)^{\maxset{0,2-\ckindsite{1}}}}{1-w t} \\
&\qquad \times \left( y^{-\frac12\ckindsite{2}} \IntOp{y}{0}{y} y^{\frac12\ckindsite{2}-1} \right) \left( y^{-\frac12\ckindsite{2}} y^{\bkind+2}\Diff{y}{\bkind+2} y^{\frac12\ckindsite{2}} \right) \\
&\qquad \times \left( y^{\frac12\ckindsite{2}-1}y^{\akind+1}\Diff{y}{\akind+1} y^{1-\frac12\ckindsite{2}} \right) y^{\frac12\ckindsite{2}} \frac{(y t)^{\maxset{0,2-\ckindsite{2}}}}{1-y t} \\
&\qquad \times  t^{\ckindsite{1}+\ckindsite{2}-3} \left(\frac{t-1}{t}\right)^{\akind+\bkind+\ckind}   \eqndot
\end{aligned}
\end{equation}
Defining an abbreviation for the second half of the respective $w$- and $y$-dependent lines, we find
\begin{equation}\label{eq:Diff_Operator_action}
\begin{aligned}
 O(x,\akind,\ckindsite{i})&=x^{\frac 12\ckindsite{i}}\left( x^{\frac 12\ckindsite{i}-1}x^{\akind+1}\Diff{x}{\akind+1} x^{1-\frac12\ckindsite{i}} \right) x^{\frac12\ckindsite{i}} \frac{(x t)^{\maxset{0,2-\ckindsite{i}}}}{1-x t} \\
&=\frac{(\akind+1)!(t x)^{\akind}}{(1- t x)^{\akind+2}}x^{\ckindsite{i}}
-\delta_{\ckindsite{i}}
\left(\delta_{\akind} +2 t x\delta_{\akind}+2 t x\delta_{(\akind-1)} \right)
-\delta_{(\ckindsite{i}-1)}\delta_{\akind}x \eqncom
\end{aligned}
\end{equation}
for $(i,x)\in\{(1,w),(2,y)\} $. This allows us to perform the sum over $\akind$ via the following identity\footnote{This identity can be found with the help of \tt{Mathematica}.}
\begin{equation} 
\begin{aligned}\label{eq:Operator_Sum_identity}
 \sum_{\akind=0}^\infty \frac{1}{\akind!(\akind+1)!} &\left(\frac{t-1}{t}\right)^{\akind} O(w,\akind,\ckindsite{1})O(y,\akind,\ckindsite{2}) \\
&=\frac{w^{\ckindsite{1}} y^{\ckindsite{2}}}{(1-t(w+y-wy))^2}\\
&\quad-\frac{y^{\ckindsite{2}}}{(1-t y)^3}
\left[
\delta_{\ckindsite{1}}\left(1+t(2w-y-2wy)\right)
-\delta_{(\ckindsite{1}-1)}w(1-ty)
\right]\\
&\quad-\frac{w^{\ckindsite{1}}}{(1-tw)^3}
\left[
\delta_{\ckindsite{2}}\left(1+t(2y-w-2wy)\right)
-\delta_{(\ckindsite{2}-1)}y(1-tw)
\right]\\
&\quad
+wy\delta_{(c_{(1)}-1)}\delta_{(c_{(2)}-1)}
+(1+2t(w+y-wy+twy))
\delta_{c_{(1)}}\delta_{c_{(2)}}\\
&\quad
+y(1+2tw)\delta_{c_{(1)}}\delta_{(c_{(2)}-1)}
+w(1+2ty)\delta_{c_{(2)}}\delta_{(c_{(1)}-1)}
\eqndot
\end{aligned}
\end{equation}
Inserting this identity into \eqref{eq:PD2_diff_Operators}, the last four lines of \eqref{eq:Operator_Sum_identity} drop out, as they are at most linear in either $w$ or $y$. This leaves us with
\begin{equation}\label{eq: PD2 int pro op 2}
\begin{aligned}
 \ev{ P \mathfrak{D}_2(w,y) }_{\text{int}}&= -2 \sum_{\ckindsite{1},\ckindsite{2},\ckind=0}^4  \sum_{\bkind=0}^\infty   G^{\gamma_i}(\ckindsite{1}, \ckindsite{2}, \ckind) \frac{1}{\bkind!(\bkind+1)!} \\
&\qquad \times  \left( w^{-\frac12\ckindsite{1}} \IntOp{w}{0}{w} w^{\frac12\ckindsite{1}-1} \right)\left( y^{-\frac12\ckindsite{2}} \IntOp{y}{0}{y} y^{\frac12\ckindsite{2}-1} \right) \\
&\qquad \times t^{\ckindsite{1}+\ckindsite{2}-3} \left(\frac{t-1}{t}\right)^{\bkind+\ckind}  w^{\bkind+2-\frac12\ckindsite{1}} y^{\bkind+2-\frac12\ckindsite{2}}  \\
&\qquad \times \Diff{w}{\bkind+2} \Diff{y}{\bkind+2}  
\frac{w^{\ckindsite{1}} y^{\ckindsite{2}}}{(1-t (w+y-wy))^2} \eqndot
\end{aligned}
\end{equation}
Writing the last factor in \eqref{eq: PD2 int pro op 2} as a power series in the variables $\hat w=w-1$ and $\hat y=y-1$ as
\begin{equation}
\begin{aligned}\label{eq: B40}
\frac{w^{\ckindsite{1}} y^{\ckindsite{2}}}{(1-t (w+y-wy))^2}=
\sum_{\alpha,\beta=0}^4 \binom{\ckindsite{1}}{\alpha}\binom{\ckindsite{2}}{\beta} 
\sum_{n=0}^\infty\frac{n+1}{(1-t)^2}
\left(\frac{t}{t-1}\right)^n 
{\hat w}^{n+\alpha} \hat{y}^{n+\beta}\eqncom
\end{aligned}
\end{equation}
we can apply the remaining derivative operators. Combining all $b$-dependent terms and substituting $l=b+2$, we finally obtain
\begin{equation}\label{eq: PD2 final}
\begin{aligned}
\ev{ P D_2(w,y) }_{\text{int}} 
&= -2 \sum_{\ckindsite{1},\ckindsite{2},\ckind=0}^4  \sum_{\alpha,\beta=0}^4  \binom{\ckindsite{1}}{\alpha}\binom{\ckindsite{2}}{\beta} G^{\gamma_i}(\ckindsite{1}, \ckindsite{2}, \ckind)  \\
&\qquad \times  \left( w^{-\frac12\ckindsite{1}} \IntOp{w}{0}{w} w^{-1} \right)\left( y^{-\frac12\ckindsite{2}} \IntOp{y}{0}{y} y^{-1} \right) \\
& \qquad \times  t^{\ckindsite{1}+\ckindsite{2}-\ckind-1} (t-1)^{\ckind-4}  (w-1)^{\alpha}(y-1)^{\beta} \\
& \qquad \times \xi_{\alpha\beta}\left(\frac{t}{t-1}(w-1) (y-1),\frac{t-1}{t}\frac{w y}{(w-1) (y-1)}\right)\eqncom
\end{aligned}
\end{equation}
where the function $\xi_{\alpha\beta}(X,Y)$ is defined as
\begin{equation}
 \xi_{\alpha \beta}(X,Y)=\sum_{l=0}^\infty  \sum_{n=0}^\infty (l-1)l^2(n+1) \binom{n+\alpha}{l} \binom{n+\beta}{l} Y^n X^l \eqndot
\end{equation}
The evaluation of this function is presented in appendix \ref{app: summation identities}. 

Anticipating the results of appendix \ref{app: summation identities}, the expression \eqref{eq: PD2 final}  does no longer contain any infinite sums. The remaining finite sums and integrals can be evaluated with the help of {\tt Mathematica}. Assembling everything in \eqref{eq:integral_representation_PD2_2}, we find the result \eqref{eq:Result_PD2}--\eqref{eq:result_functions_last}.

\section{The calculation of \texorpdfstring{$Z_{\text{f.s.c.}}^{(1)}(x)$ 
}{Z(f.s.c.)(x)}}\label{app: corrections}
In this appendix, we compute the $L=2$ finite-size corrections to the single-trace partition function of the $\beta$-deformation with gauge group \SUN. They arise when using the finite-size-corrected dilatation-operator density \eqref{eq:D2_beta} instead of the asymptotic version \eqref{eq: asymptotic dila}. The calculation of $\ev{P\mathfrak{D}^{L=2}_2(w,y)}$ differs from that of $\ev{P\mathfrak{D}^{L\geq3}_2(w,y)}$ only in the sums over fermionic oscillators. Hence, it is sufficient to give the appropriate definition of the fermionic occupation number function $G^\beta_{L=2}(\ckindsite1,\ckindsite2,\ckind)$, which has to replace the asymptotic function $G^{\gamma_i}(\ckindsite1,\ckindsite2,\ckind)$ in the derivation of appendix \ref{app: PD2 calculation}.

According to the prescription \eqref{eq:D2_beta} for the finite-size-corrected dilatation-operator density $\mathfrak{D}^{L=2}_2$, the deformation parameter $\beta$ in $\mathfrak{D}^{L\geq3}$ has to be set to zero whenever the fields $A_{i}$ at site $1$ and $2$ are either taken from the subalphabet $\mathcal{A}_{\mathrm{matter}}$ or from $\bar{\mathcal{A}}_{\mathrm{matter}}$, which were defined in \eqref{eq:matter_type}. In the oscillator picture, these restrictions translate to the constraints
\begin{equation}
\begin{aligned}
A_i\in \mathcal{A}_{\mathrm{matter}}\Leftrightarrow \sum_{e=1}^3c_{(i)}^e=1\eqncom
\qquad\qquad
A_i\in \bar{\mathcal{A}}_{\mathrm{matter}}	\Leftrightarrow \sum_{e=1}^3c_{(i)}^e=2\eqncom
\end{aligned}
\end{equation}
which have to be included in the fermionic occupation number function $G^{\gamma_i}(\ckindsite1,\ckindsite2,\ckind)$ of \eqref{eq: def G(c1,c2,c)}.
This yields
\begin{equation} 
\begin{aligned}
 G_{L=2}^{\beta}(\ckindsite1,\ckindsite2,\ckind)&= \prod_{e=1}^4\bigg(\sum_{\ckindsite[e]{1},\ckindsite[e]{2},\ckind[e]=0}^1 \binom{\ckindsite[e]{1}}{{\ckind[e]}}\binom{\ckindsite[e]{2}}{\ckind[e]}(-1)^{\ckind[e]}\bigg) \\
 &\qquad \times \e^{-\complexi \sum_{l,m=1}^4 \ckindsite[l]{1}\ckindsite[m]{2}\bq_{\psi^{l}}\times\bq_{\psi^{m}}}\Big|_{
 	\substack{
 		\beta=0 \text{ if } \sum_{e=1}^3c_{(1)}^e=\sum_{e=1}^3c_{(2)}^e=1 \\
 		\text{or if } \sum_{e=1}^3c_{(1)}^e=\sum_{e=1}^3c_{(2)}^e=2}
 	} \\
 &\qquad \times \delta_{\left(\ckindsite{1}-\sum_{e=1}^4\ckindsite[e]{1}\right)}
 \delta_{\left(\ckindsite{2}-\sum_{e=1}^4\ckindsite[e]{2}\right)}
 \delta_{\left(\ckind-\sum_{e=1}^4\ckind[e]\right)}\eqndot
\end{aligned}
\end{equation}
Inserting $\G_{L=2}^{\beta}(\ckindsite1,\ckindsite2,\ckind)$ into \eqref{eq: PD2 pre dual way} and following the derivation of appendix \ref{app: PD2 calculation} yields $\ev{P\mathfrak{D}^{L=2}_2(w,y)}$. Combining it with the asymptotic expression $\ev{P\mathfrak{D}^{L\geq3}_2(w,y)}$, we obtain the finite-size correction \eqref{eq: Z_corr in full theory}.


\section{Summation identities}\label{app: summation identities}
In this appendix, we derive summation identities for
\begin{equation}
 \xi_{\alpha \beta}(X,Y)=\sum_{j=0}^\infty  \sum_{i=0}^\infty (i-1)i^2(j+1) \binom{j+\alpha}{i} \binom{j+\beta}{i} Y^i X^j \eqncom
\end{equation}
with $\alpha, \beta = 0,1,2,3,4$. This can be achieved by applying a finite number of derivative and integral operators and using the following identities:
\begin{equation}\label{eq:P_identity}
  \sum_{i=0}^j \binom{j}{i}^2 t^i=(1-t)^j P_j\left(\frac{1+t}{1-t}\right)\eqncom\qquad
 \sum_{n=0}^\infty P_n(x) z^n =  (1 - 2xz + z^2)^{-1/2} \eqncom
\end{equation}
where $P_n(x)$ denotes the $n^{\text{th}}$ Legendre polynomial.

Note that $\xi_{\alpha \beta}$ is symmetric in $\alpha$ and $\beta$. Assuming that $\alpha\geq\beta$, we find
\begin{equation} 
\begin{aligned}\label{eq: xi in operators}
& (i-1)i^2(j+1)\binom{j+\alpha}{i} \binom{j+\beta}{i} Y^i X^j \\
&= (i-1)i^2(j+1)  \prod_{\gamma=\beta+1}^{\alpha}\left(1-\frac{i}{j+\gamma}\right) \binom{j+\alpha}{i}^2 Y^i X^j \\
&= \left(Y^2\Diff{Y}{2}\right)\left(Y\Diff{Y}{}\right)\left(\Diff{X}{} X\right) \\ &\qquad\qquad\prod_{\gamma=\beta+1}^{\alpha}\left[1-\left(Y\Diff{Y}{}\right) \left(\frac{1}{X^\gamma} \IntOp{X}{0}{X} X^{\gamma-1}\right)\right] \binom{j+\alpha}{i}^2 Y^i X^j \eqndot
\end{aligned}
\end{equation}
Using \eqref{eq:P_identity}, we have
\begin{equation} 
\begin{aligned}\label{eq: xi Legendre}
 &\sum_{i,j=0}^{\infty} \binom{j+\alpha}{i}^2 Y^i X^j 
= \frac{1}{X^\alpha}\sum_{j=0}^{\infty}  (X(1-Y))^{j+\alpha}P_{j+\alpha}\left(\frac{1+Y}{1-Y}\right) \\
&= \frac{1}{X^\alpha} \left[ \frac{1}{\left(1-2X(1+Y)+X^2(1-Y)^2\right)^{1/2}}-\sum_{k=0}^{\alpha-1}  (X(1-Y))^k P_k\left(\frac{1+Y}{1-Y}\right) \right] \eqndot
\end{aligned}
\end{equation}
Thus, combining \eqref{eq: xi in operators} with \eqref{eq: xi Legendre} allows to express $\xi_{\alpha \beta}(X,Y)$ in a form explicitly solvable with {\tt{Mathematica}}:
\begin{equation} 
\begin{aligned}
\xi_{\alpha \beta}(X,Y)&=
\left(Y^2\Diff{Y}{2}\right)\left(Y\Diff{Y}{}\right)\left(\Diff{X}{} X\right) \prod_{\gamma=\beta+1}^{\alpha}\left(1-\left(Y\Diff{Y}{}\right) \left(\frac{1}{X^\gamma} \IntOp{X}{0}{X} X^{\gamma-1}\right)\right) 
\\
&\phaneq \frac{1}{X^\alpha} \left[ \frac{1}{\left(1-2X(1+Y)+X^2(1-Y)^2\right)^{1/2}}-\sum_{k=0}^{\alpha-1}  (X(1-Y))^k P_k\left(\frac{1+Y}{1-Y}\right) \right] \eqndot \\
\end{aligned}
\end{equation}
For example for $(\alpha,\beta)=(2,1)$, we obtain
\begin{equation} 
\begin{aligned}
\xi_{2 1}\left(X,Y\right)=& \frac{24 X Y^2}{\left(1-2 X (1+Y)+X^2 (1-Y)^2\right)^{9/2}}\\
&{}\Bigl[{}1-X^2(5-12Y+5Y^2)
+X^3(5-6Y-6Y^2+5Y^3)\\
&-9 X^4(1-Y)^2Y
-X^5(1-Y)^4(1+Y)\Bigr]
\end{aligned}
\end{equation}
The remaining expressions give no further insights, and hence we refrain from showing them.


\section{Path-integral calculation} \label{app: path integral}
In this appendix, we provide a non-trivial test of our \Polya-theoretical approach by comparing the one-loop partition function of the deformed \NfSYMt with gauge group \UN with the one obtained from a Feynman-diagrammatic approach. We briefly present the results of the latter approach for the undeformed \NfSYMt \cite{Mussel:2009uw} and provide the necessary modifications for $\beta$- and \gidefd theories with gauge group \UN. We find agreement for both approaches for gauge group \UN and comment on the generalisation for gauge group \SUN.

In \cite{Mussel:2009uw}, a Feynman-diagrammatic approach\footnote{The underlying Feynman-diagrammatic formalism for gauge theories on \RxSt was developed in \cite{AMMPR03,AMMPR05,AMR06}.} was used to calculate the one-loop partition function\footnote{Note that our notion of loops refers to divergent loops in two-point functions or powers of $g^2$, whereas the notion of loops in \cite{AMMPR03,AMMPR05,AMR06,Mussel:2009uw} refers to loops in vacuum bubbles. As a consequence, `two-loop' in the counting of \cite{Mussel:2009uw} corresponds to `one-loop' in our counting.} of a class of Yang-Mills theories on \RxSt with matter fields transforming in the adjoint representation of the gauge group \UN. This calculation employs the vacuum path integral of the Eucidean theory on $\text{S}^1\times \text{S}^3$, where the circumference of $\text{S}^1$ is $1/RT=-\ln x$. The Euclidean action\footnote{Note that the normalisation of the coupling constant $\tilde{g}_\YM$ of \cite{Mussel:2009uw} differs by a factor of $\sqrt{2}$ from our $g_\YM$.} with $N_s$ scalars $\Phi^a$, $a=1,\dots,N_s$, and $N_f$ fermions $\Psi^I$, $I=1,\dots,N_f$ is assumed to take the form
\begin{equation}\label{eq: FlatAction}
\begin{aligned}
S_E &= \int_0^{\frac{1}{RT}}\measure{t} \int_{\text{S}^3}\measure{\Omega} \, \tr \Big\{\frac{1}{4}F_{\mu\nu}F^{\mu\nu}+\frac{1}{2}\Phi^a\left(-\D^2+1\right)\Phi^a+\complexi\Psi^{\dag I}\sigma^{\mu}\D_{\mu}\Psi_I \\
&\phaneq -\frac{1}{4}\tilde{g}_\YM^2\mathbf{Q}^{abcd}\Phi_a\Phi_b\Phi_c\Phi_d+ \frac{1}{2}\tilde{g}_\YM(\rho^{a\dag})^{IJ}\Psi_{I}\varepsilon[\Phi^a,\Psi_J]+\frac{1}{2}\tilde{g}_\YM\rho^a_{IJ}\Psi^{\dag I}\varepsilon[\Phi^a,\Psi^{\dag J}]\Big\}\eqncom
\end{aligned}
\end{equation}
where $\varepsilon$ is used to contract the spinor indices and the coupling tensors $\mathbf{Q}^{abcd}$ and $\rho^a_{IJ}$ are general but assumed to stem from commutator interactions. After gauge fixing and integrating out the non-zero modes in the path integral, one obtains an effective action depending on the effective remaining zero-mode $U$ and $x=\e^{-1/RT}$. Up to order $\tilde{g}_\YM^2$, it has the form $S_{\text{eff}}(U,x)=S_{\text{eff}}^{\text{1-loop}}+S_{\text{eff}}^{\text{2-loop}}$, with
\begin{align}
S_{\text{eff}}^{\text{1-loop}} &= -\sum_{n=1}^{\infty}\frac{1}{n}z_n(x)\tr(U^n)\tr(U^{\dag n})\eqncom\label{eq: 1LoopEffAction}\\
S_{\text{eff}}^{\text{2-loop}} &= -\tilde{g}_\YM^2\ln x \bigg[N\sum_{n=1}^\infty f_n(x)\left(\tr(U^n)\tr(U^{\dag n})-1\right)  \nnl
&\phaneq\phantom{\beta g^2 [}+\sum_{n, m=1}^{\infty}f_{n m}(x)\left(\tr(U^n)\tr(U^m)\tr(U^{-n-m}) + c.c.-2N\right)\bigg] \eqndot
\label{eq: 2LoopEffAction}
\end{align}
The coefficients $f_n(x)$ and $f_{n m}(x)$ have to be calculated via Feynman diagrams and $z_n(x)= z(\omega^{n+1}x^n)$ has the same interpretation as the single-site partition function \eqref{eq: single site partition function}. 

In the large $N$ limit and for a temperature below the critical temperature, the Euclidean path-integral can be solved in the saddle point approximation; see \cite{AMMPR03} for details. For the \UN theory, the final result for the multi-trace partition function is 
\begin{equation}\label{eq: UnZ}
\cZ_{\UN}(x)=
x^{\tilde\lambda \tilde{F}_2^{np}(x)}\prod_{n=1}^{\infty}\frac{x^{-\tilde\lambda f_n(x)}}{1-z_n(x)-\tilde\lambda n f_n(x)\ln x }\eqncom
\end{equation}
where $\widetilde{F}_2^{np}(x)= -2\sum_{n,m=1}^{\infty}f_{nm}(x)$ and the \thooft coupling takes the form $\tilde\lambda=N\tilde{g}_\YM^2$. The explicit Feynman-diagram calculation of \cite{Mussel:2009uw} yields the coefficients:
\begin{align}
f_n(x)&=f_{1,+}(x^n)+(-1)^{n+1}f_{1,-}(x^n) \eqncom\\
\intertext{with}
f_{1,+}(x)&=\frac{1}{\pi^2}\left[\left(-\frac{3}{8}N_s + \frac{1}{8}\tr(\rho^{a\dag}\rho^a)\right)x\right.\nnl*
&\phaneq+\left(\frac{1}{4}+\frac{1}{2}N_f+\frac{1}{8}N_s-\frac{1}{8}\mathbf{Q}^{abba}+\frac{1}{2}\tr(\rho^{a\dag}\rho^a)\right)x^2 \nnl*
&\phaneq+\left(4+2N_f+\frac{25}{8}N_s-\frac{1}{2}\mathbf{Q}^{abba}+\frac{11}{8}\tr(\rho^{a\dag }\rho^a)\right)x^3 \nnl*
&\phaneq+\left(\frac{55}{4}+5N_f+\frac{41}{4}N_s-\frac{5}{4}\mathbf{Q}^{abba}+3\tr(\rho^{a\dag}\rho^a)\right)x^4 \nnl*
&\phaneq+\left(32+10N_f+\frac{185}{8}N_s-\frac{5}{2}\mathbf{Q}^{abba}+\frac{45}{8}\tr(\rho^{a\dag}\rho^a)\right)x^5 \nnl*
&\phaneq\left.+\left(\frac{245}{4}+\frac{35}{2}N_f+\frac{347}{8} N_s-\frac{35}{8}\mathbf{Q}^{abba}+\frac{19}{2}\tr(\rho^{a\dag}\rho^a)\right)x^6+\cdots\right]\eqncom\\
f_{1,-}(x) &=\frac{1}{\pi^2}\left[\frac{1}{4}\tr(\rho^{a\dag}\rho^a) x^{\frac{3}{2}} + (3 N_f +\frac{5}{4} \tr(\rho^{a\dag}\rho^a)) x^{\frac{
		5}{2}} \right. \nnl*
&\phaneq + (12 N_f +\frac{7}{2}\tr(\rho^{a\dag}\rho^a)) x^{\frac{7}{2}} + (30N_f +\frac{15}{2} \tr(\rho^{a\dag}\rho^a)) x^{\frac{9}{2}}  \nnl*
&\phaneq\left.+(60N_f +\frac{55}{4}\tr(\rho^{a\dag}\rho^a)) x^{\frac{11}{2}}+\cdots\right]  \eqncom \\
\widetilde{F}_2^{np}(x) &=\frac{\mathbf{Q}^{aabb}}{8\pi^2}x^2-\frac{\tr(\rho^{a\dag }\rho^a)}{\pi^2}x^{\frac{5}{2}}\nnl*
&\phaneq+\frac{\left[4\mathbf{Q}^{aabb}\!-\!2\tr(\rho^{a\dag }\rho^a)\!-\!6N_f\!-\!9N_s\right]}{4\pi^2}x^3\!-\!\frac{4\tr(\rho^{a\dag }\rho^a)}{\pi^2}x^{\frac{7}{2}} \nnl*
&\phaneq+\frac{18\!+\!23\mathbf{Q}^{aabb}\!-\!48N_s}{8\pi^2}x^4\!+\!\frac{3(N_f\!-3\tr(\rho^{a\dag }\rho^a))}{\pi^2}x^{\frac{9}{2}} \nnl*
&\phaneq+\!\frac{\!14\mathbf{Q}^{aabb}\!-\!\tr(\rho^{a\dag }\rho^a)\!-9N_f-\!39N_s}{2\pi^2}x^5-\frac{20\tr(\rho^{a\dag }\rho^a)}{\pi^2}x^{\frac{11}{2}} \nnl*
&\phaneq+\frac{42+50\mathbf{Q}^{aabb}-6\tr(\rho^{a\dag }\rho^a)-18N_f-117N_s}{4\pi^2}x^6+\cdots \eqncom
\end{align}
where we have to set $N_f=4$ and $N_s=6$ to obtain the result for \NfSYMt and its deformations. With these findings, the authors of \cite{Mussel:2009uw} were able to reproduce the result derived from the \Polya-theoretic approach \cite{Spradlin:2004pp} for the undeformed \NfSYMt.

The result for the multi-trace partition function \eqref{eq: UnZ} can be modified\footnote{We thank Ofer Aharony and Ran Yacobi for a very helpful discussion on the exact nature of the necessary modifications.} to also capture the $\beta$- and \gidef of \NfSYM, which have non-commutator interactions. To this end, the simplifications which are only justified for commutator interactions have to be undone. In \eqref{eq: 2LoopEffAction}, the terms which are independent of the zero-mode $U$ stem from non-planar contributions. For non-commutator interactions, they have coefficients $\hat f_n(x)$ and $\hat f_{nm}(x)$ which are independent of the respective planar coefficients. Hence, we have to generalise \eqref{eq: 2LoopEffAction} to
\begin{equation}\label{eq: 2LoopEffActionmodified}
\begin{aligned}
S_{\text{eff}}^{\text{2-loop}} &= - \tilde{g}_\YM^2\ln x\bigg[N\sum_{n=1}^\infty\left( f_n(x)\tr(U^n)\tr(U^{\dag n})-\hat{f}_n(x)\right) \\
&\phaneq\phantom{\beta g^2[}+\sum_{n,m=1}^{\infty}\left(f_{n m}(x)\left(\tr(U^n)\tr(U^m)\tr(U^{-n-m}) + c.c.\right)-2N\hat{f}_{n m}(x)\right)\bigg] \eqndot
\end{aligned}
\end{equation} 
Defining $\hat{\tilde{F}}_2^{np}(x)= -2\sum_{n,m=1}^{\infty}\hat{f}_{nm}(x)$ and performing the saddle point approximation then yields
\begin{equation}\label{eq: UnZmodified}
\begin{aligned}
\cZ_{U(N)}(x)=x^{\tilde\lambda \hat{\tilde{F}}_2^{np}(x)}\prod_{n=1}^{\infty}\frac{x^{-\tilde\lambda \hat{f}_n(x)}}{1-z_n(x)\tilde\lambda n f_n(x)\ln x}\eqndot
\end{aligned}
\end{equation}
While the planar contributions $f_n(x)$ and $f_{n m}(x)$ only contain the contracted coupling tensors $\tr(\rho^{a\dag}\rho^a)$ and $\mathbf{Q}^{aabb}=\mathbf{Q}^{abba}$, the non-planar contributions $\hat{f}_n(x)$ and $\hat{f}_{n m}(x)$ must contain $\tr(\rho^{a*}\rho^{a})$ and $\mathbf{Q}^{abab}$. In the case of commutator interactions, the non-planar contributions can be expressed in terms of planar ones via $\rho^{a\dag}=(\rho^{aT})^*=-\rho^{a*}$ and $\mathbf{Q}^{abab}=-2 \mathbf{Q}^{abba}$.\footnote{Mind the typo in footnote 21 of \cite{Mussel:2009uw}.} Thus, to reconstruct $\hat{f}_n(x)$ and $\hat{f}_{n m}(x)$ from $f_n(x)$ and $f_{n m}(x)$, we have to make the following replacements:
\begin{equation}
\begin{gathered}
\tr(\rho^{a\dag}\rho^a) \rightarrow -\tr(\rho^{a*}\rho^a) \eqncom\\
\mathbf{Q}^{abba} \rightarrow -\frac12 \mathbf{Q}^{abab} \eqncom \qquad \mathbf{Q}^{aabb} \rightarrow -\frac12 \mathbf{Q}^{abab} \eqndot
\end{gathered}
\end{equation}
The coupling tensors $\mathbf{Q}^{abcd}$ and $\rho^a_{IJ}$ can be determined by comparing the action of $\beta$- and \gidefd \NfSYMt with gauge group \UN \eqref{eq: component action single trace} with the action in \eqref{eq: FlatAction}. To this end, we expand the complex scalars in \eqref{eq: component action single trace} as $\phi^i=(\Phi^i+\complexi\Phi^{i+3})/\sqrt{2}$ and cyclically symmetrise the fields in the quartic scalar vertex. Recall that the coupling constant is $\tilde{g}_\YM=g_\YM/\sqrt{2}$, and that the action \eqref{eq: FlatAction} appears as $\e^{-S}$ in the path integral, whereas the action \eqref{eq: component action single trace} appears as $\e^{S}$ in the path integral. The relevant contractions of the coupling tensors are
\begin{equation} 
\begin{aligned}
\mathbf{Q}^{aabb}=\mathbf{Q}^{abba}=-30 \eqncom&\qquad \mathbf{Q}^{abab}= 60-24\sum_{i=1}^3\sin^2\frac{\gamma_i^++\gamma_i^-}{2} \eqncom&\\
\tr(\rho^{a\dag}\rho^a)= 24 \eqncom&\qquad
\tr(\rho^{a*}\rho^a)= 24-8\sum_{i=1}^3\left(\sin^2\frac{\gamma_i^+}{2}+\sin^2\frac{\gamma_i^-}{2}\right) \eqndot
\end{aligned}
\end{equation}
The power series obtained from \eqref{eq: UnZmodified} for the multi-trace partition function confirms our result for gauge group \UN.\footnote{We verified this up to order 6 in $x$, which is the maximum order given in \cite{Mussel:2009uw}.}

Generalising the above analysis to the $\beta$- and \gidef with gauge group \SUN is considerably harder, since the calculation of \cite{Mussel:2009uw} manifestly uses the Feynman rules derived for gauge group \UN. For commutator interactions, the \U1 fields are free and their contribution to the partition function can simply be divided out in the final result. This is no longer true in the case of non-commutator interactions; in the $\beta$- and \gidef the \U1  matter modes couple to the \SUN matter modes as explicitly shown in \cite{Fokken:2013aea}. Therefore, to obtain the one-loop partition function in the \SUN case, modifications at the level of Feynman diagrams would be necessary. These modifications would have to include the \SUN propagators of all fields and in case of the $\beta$-deformation the contributions from the double-trace coupling \eqref{eq: component action double trace}.

\phantomsection





\bibliographystyle{JHEP}
\bibliography{ThesisINSPIRE}
%

\end{document}